\documentclass[pra,superscriptaddress,twocolumn]{revtex4}

\usepackage{amsmath}
\usepackage{amsfonts}       
\usepackage{graphicx}
\usepackage{dcolumn}
\usepackage{epsfig}
\usepackage{bm}
\usepackage{array}
\usepackage{hyperref}
\hypersetup{
colorlinks=true,
citecolor=blue,
linkcolor=blue,
urlcolor=blue,
}
\usepackage{xcolor}			
\usepackage{graphicx}
\usepackage{amsmath}
\usepackage{amssymb}
\usepackage{bm}           
\usepackage{bbm}
\usepackage[frenchb]{babel}
\usepackage{ulem}
\usepackage{setspace}
\usepackage{wasysym}
\usepackage{adjustbox}
\usepackage{framed}
\usepackage{mdframed}
\usepackage{latexsym} 
\usepackage{pdflscape}   
\usepackage{multirow}    
\usepackage{booktabs}    
\usepackage{calc}        

\usepackage{calrsfs}								
\DeclareMathAlphabet{\pazocal}{OMS}{zplm}{m}{n}	

\definecolor{DeepSkyBlue}{RGB}{0,104,139}
\colorlet{MySky}{white!40!blue}
\colorlet{MyViolet}{red!45!blue}
\colorlet{MyBlue}{black!40!blue}
\colorlet{MyRed}{black!40!red}
\colorlet{MyOrange}{red!70!yellow}
\colorlet{MyGreen}{black!60!green}
\colorlet{MyBrown}{black!70!brown}
\colorlet{MyGray}{black!60!white}

\newcommand{\be}{\begin{equation}}
\newcommand{\ee}{\end{equation}}

\newcommand{\beq}{\begin{eqnarray}}
\newcommand{\eeq}{\end{eqnarray}}

\newcommand{\bra}[1]{\ensuremath{\langle#1|}}
\newcommand{\ket}[1]{\ensuremath{|#1\rangle}}

\UseRawInputEncoding

\begin{document}

\title{Can multipartite entanglement be characterized\\ by two-point connected correlation functions ?}

\author{Luca Lepori}
\email[correspondence at: ]{luca.lepori@unipr.it}
\affiliation{Dipartimento di Scienze Matematiche, Fisiche e Informatiche, Universit\`a  di Parma, Parco Area delle Scienze, 53/A, I-43124 Parma, Italy.}
\affiliation{INFN, Gruppo Collegato di Parma, Parco Area delle Scienze 7/A, 43124, Parma, Italy.}
\affiliation{QSTAR and INO-CNR, Largo Enrico Fermi 2, 50125 Firenze, Italy.}

\author{Andrea Trombettoni}
\affiliation{Dipartimento di Fisica, Universit\`a di Trieste, Strada Costiera 11, I-34151Trieste, Italy.}
\affiliation{CNR-IOM DEMOCRITOS Simulation Center and SISSA, Via Bonomea 265,I-34136 Trieste, Italy.}

\author{Domenico Giuliano}
\affiliation{Dipartimento di Fisica, Universit\`a della Calabria, Arcavacata di Rende I-87036, Cosenza, Italy.}
\affiliation{INFN, Gruppo collegato di Cosenza, Arcavacata di Rende I-87036, Cosenza, Italy.}

\author{Johannes Kombe}
\affiliation{Department of Physics and SUPA, University of Strathclyde, Glasgow G4 0NG, United Kingdom.}

\author{Jorge Yago Malo} 
\affiliation{Dipartimento  di  Dipartimento di Ricerca traslazionale e delle nuove tecnologie in Medicina e Chirurgia, via Savi 10, 56126 Pisa.}
\affiliation{Dipartimento  di  Fisica  Enrico  Fermi, Universit\`a  di  Pisa  and  INFN, Largo  B.  Pontecorvo  3,  I-56127  Pisa,  Italy.}

\author{Andrew J. Daley}
\affiliation{Department of Physics and SUPA, University of Strathclyde, Glasgow G4 0NG, United Kingdom.}

\author{Augusto Smerzi}
\affiliation{QSTAR and INO-CNR, Largo Enrico Fermi 2, 50125 Firenze, Italy.} 
\affiliation{European Laboratory for Nonlinear Spectroscopy (LENS), Universit\`a  di Firenze, 50019 Sesto Fiorentino, Italy.}

\author{Maria Luisa Chiofalo}
\affiliation{Dipartimento  di  Fisica  Enrico  Fermi, Universit\`a  di  Pisa  and  INFN, Largo  B.  Pontecorvo  3,  I-56127  Pisa,  Italy.}

\begin{abstract}
 We discuss under which conditions multipartite entanglement in
 mixed quantum states can be characterized only in terms of two-point connected correlation functions, as it is the case for pure states.
{\color{black} In turn, the latter correlations are defined via a suitable combination of (disconnected) one- and two-point correlation functions.}
In contrast to the case of pure states,   
conditions to be satisfied turn out to be rather severe.  However, we were able to identify some interesting cases, as when 
 the point-independence is valid of the one-point correlations 
in each possible decomposition of the density matrix, or when  the operators that enter in the correlations are (semi-)positive/negative defined.
\end{abstract}

\maketitle

\section{Introduction} \color{black}
{\color{black} Entanglement is a central resource in quantum technology}~\cite{AmicoRMP2008, PezzeRMP,toth2014}. Entanglement quantifiers are especially relevant in {\color{black} quantum information,} quantum metrology and estimation theory~\cite{toth2009,maccone2011,pezze2014,huber2019,HyllusPRA2012,TothPRA2012,smerzi2018}, {\color{black} and are recognized as valuable tools to describe quantum phases and phase transitions~\cite{HorodeckiRMP2009, toth2009}.
Research has largely} focused on bipartite entanglement~\cite{AmicoRMP2008, eisert2010,Zeng}, 
typically via the von Neumann entropy~\cite{AmicoRMP2008, Zeng, LatorreJPA2009, eisert2010, OsbornePRA2002, VidalPRL2003},
entanglement spectrum~\cite{LiPRL2008,PollmannPRB2010,FidkowskiPRL2010,ThomalePRL2010,lepori2012,leporiLR}, and generally pairwise entanglement~\cite{OsterlohNATURE2002,LidarPRL2004,AmicoRMP2008,eisert2010,MartyPRL2016}.\\
\indent 
 More recently, considerable attention has been devoted to the so-called multipartite entanglement (ME) among subsets of a quantum state, a concept based on those of producibility and separability~\cite{pezze2014}. However, the identification and detection of ME criteria are still a challenging problem.
Prominent tools for ME quantification are the quantum Fisher information (QFI), presently known as the ultimate bound for ME \cite{PezzePRL2009, HyllusPRA2012,TothPRA2012,pezze2014,gessner2016,ren2021},
and Wineland's spin-squeezing parameter (SSP)~\cite{WinelandPRA1994,MaPHYSREP2011,PezzeRMP,sorensen2000,molmer2001}. 
 Methods to extract them from experiments 
have been discussed 
~\cite{hauke2016,PezzePNAS2016,WinelandPRA1994,MaPHYSREP2011,PezzeRMP,molmer2001}, 
and results for ultracold atomic gases have been  presented~\cite{sorensen2000,StrobelSCIENCE2014}.
QFI and SSP also provide efficient tomography~\cite{pezze2018,mirkhalaf2020} for quantum phases and transitions~\cite{MaPRA2009, LiuJPA2013, hauke2016, pezze2017, ZhangPRL2018, gabbrielli2018,lucchesi2019}, and are proposed as benchmarks for quantum simulators \cite{pezze2018}. In fact, ME has been investigated {\color{black} e.g. in Bose-Einstein condensates \cite{sorensen2000,StrobelSCIENCE2014},  spin systems~\cite{GuhneNJP2005,hauke2016, LiuJPA2013, MaPRA2009,lucchesi2019,daley2020}, also at criticality~\cite{HofmannPRB2014,calabrese2004,roscilde2018,roscilde2019}, with long-range interactions~\cite{koffel2012,lepori2016,vodola2016,ares2015,lepori2018}, and in topological models~\cite{orus2014,pezze2017,ZhangPRL2018,preskill2006,nori2021}.\\
\indent Estimating multipartite entanglement for mixed states via the QFI 
can be difficult in general, and in strongly-interacting systems it has been performed only in a limited number of cases, mainly at  equilibrium~\cite{gabbrielli2018,daley2020},   also establishing a direct link with dynamical susceptibility  \cite{hauke2016}. The main reason is that the QFI {\color{black} for a mixed state cannot be fully expressed in terms of (one- and two-point) correlation functions, instead it requires a sum over matrix elements with the respect to all the states diagonalizing the full density matrix~\cite{gabbrielli2018} (see the next Section for a precise definition).} \\
\indent In this work, focusing on 
 mixed quantum states, we discuss {\color{black} physical and mathematical conditions that make it possible} to bound
 multipartite entanglement, via 
 one- and two-point correlation functions.
 The conditions to be satisfied turn out to be rather severe.  However, we were able to identity {\color{black} at least two physically relevant situations,} such as when 
 the point-independence is valid of the one-point correlations 
in each possible decomposition of the density matrix or when  the operators that enter in the correlations are (semi-)positive/negative defined.
  
\section{Multipartite Entanglement}

\subsection{Separability and producibility}
\label{sep}
{\color{black} For a  $d$-dimensional discrete system with $N$ components (e.g., sites), $c$-partite entanglement, with $1\leq c \leq N$,} implies that a partition $\{ \ket{\psi_i} \}$ exists, where the maximum number of components {\color{black} in a single} $\ket{\psi_i}$ is $c$. {\color{black} The tensor-product state $\ket{\psi} = \prod_{\otimes_i} \,  \ket{\psi_i}$} is then said to be $c$-producible~\cite{pezze2014}, or to have entanglement depth $c$.  {\color{black} In addition, a system is said to host $c$-partite entanglement if it is $c$-producible but not $(c+1)$-producible. Instead, the} number $h$, with $ \frac{N}{c}   \leq h \leq N - c +1$, of disentangled subsets is the degree of separability \cite{toth2009,ren2021}.
The usual separability corresponds to $h=N$ and $c=1$. On a lattice, the subsystems are not necessarily adjacent sites. 
When $c = N$, $\ket{\psi}$ is  said to host {\it genuine} ME \cite{notec}. \\
\indent  For mixed states, {\color{black} $c$-producibility in $h$ subsets} holds if  $\rho$ can be decomposed (generally not uniquely) as 
\beq
\rho =
  \sum_{\tilde{\lambda}} p_{\tilde{\lambda}} \,  \ket{\tilde{\lambda}} \bra{\tilde{\lambda}}  \, ,
\label{cpartmix31} 
\eeq
{\color{black} where $ p_{\tilde{\lambda}}>0$  without any lack of generality, and $\ket{\tilde{\lambda}}$ are c-separable states in $h$ subsets, not necessarily with the same space-partition.} 
 If $c=N$, then Eq. \eqref{cpartmix31} is still valid, trivially with a single partition and in every decomposition.
In general, the $c$-producible decomposition $\ket{\tilde{\lambda}}$ is not orthogonal, thus $\rho$ is not diagonal. 
Moreover, the producibility of $\ket{\tilde{\lambda}}$ is generally lost in other decompositions.

\subsection{Bounds for multipartite entanglement}
\label{bme}

First, we focus on pure states $| \psi \rangle$. {\color{black} We denote by $x,y$ two sites of the lattices and by $\hat{o}(x)$, $\hat{o}(y)$ local operators based on these sites. In this way, the variance of the Hermitian operator
\beq
\hat{O} = \sum_x \hat{o}(x)
\eeq
on $| \psi \rangle$ is defined as \cite{pezze2014}}
\beq
V[\ket{\psi}, \hat{O}]_N = 4\sum_{x,y}\langle \psi | \hat{o} (x) \hat{o} (y) | \psi \rangle_{\mathrm{c}} \, ,
\label{qidef}
\eeq
being
\beq
\langle \psi | \hat{o} (x) \hat{o} (y) | \psi \rangle_{\mathrm{c}} \equiv  \langle \psi | \hat{o} (x) \hat{o} (y) | \psi \rangle -\langle \psi | \hat{o} (x) | \psi \rangle\langle \psi | \hat{o} (y) | \psi \rangle
\eeq
 two-point {\it connected} correlations~\cite{noteconv}.  $V[\ket{\psi}, \hat{O}]_N$ coincides with the QFI  for pure states \cite{pezze2014}. 
 {\color{black}  In the following, the QFI for pure states will be also denoted as $F[\ket{\psi} , \hat{O}]_N$.
Importantly, the same quantity is a witness of ME for these states. This means that, if $(c,h)$-entanglement, but not $(c+1, h^{\prime})$-entanglement (with $h^{\prime} \geq \frac{N}{c+1}$), is present, then the inequality
\be 
 V[\ket{\psi}, \hat{O}]_N \leq 4 \, k \, \big[ c \, (N - h) + N \big] \, ,
\label{FQineq}
\ee
holds \cite{ren2021}.} Indeed, \eqref{FQineq} bounds the quantum advantage offered by $(c,h)$-entanglement in terms of the sensitivity, with respect to the shot-noise (separable) limit, $ V[\ket{\psi}, \hat{O}]_N = 4 \, k \, N$. In \eqref{FQineq}, $k = \frac{(m - n)^2}{4}$ if  $\hat{o} (x)$ is constrained, with eigenvalues {\color{black} $n \leq  q  \leq m < \infty$} \cite{noteconv}.
More in detail,  \eqref{FQineq} generates other relevant bounds \cite{ren2021}: for instance, it can be extended choosing $h =  \frac{N}{c} $,
which yields the bound  for $c$-producibility $V[\ket{\psi}, \hat{O}]_N \leq 4 \, k \, c \, N $~\cite{PezzePRL2009,pezze2014}.
This implies that its violation signals at least $(c+1)$-partite entanglement \cite{HyllusPRA2012,TothPRA2012}.
The ultimate limit  $V[\vert \psi \rangle, \hat{O}]_N \, = 4 \, k \, N^2$, when $\ket{\psi}$ hosts genuine ME, is called the Heisenberg limit.
Another similar estimator for $c$, generally supposed to be not an integer, was found in ~\cite{gessner2016}, see SM 1. Similarly, one can maximize in $c$ the right-hand term of Eq. \eqref{FQineq}, setting $c = N - h + 1$ and obtaining the bound for $h$:
$V[\ket{\psi}, \hat{O}]_N \leq 4 \, k \, \Big[(N-h+1)^2 + h - 1\Big]$ \cite{song2015,ren2021}.

Finally, we notice that, critically, if two lattice sites $x$ and $y$ belong to different partitions and $\ket{\psi}$ is producible, then $\langle \psi | \hat{o} (x) \hat{o} (y) | \psi \rangle_{\mathrm{c}} = 0$, for every conceivable local operator $\hat{o}$, see e. g. \cite{shi2003,shi2004}. {\color{black} This property
can be exploited to demonstrate the bounds of producibility for pure states recalled above, see SM 2.}

Now, we focus on mixed states. 
{\color{black} We define as 
\beq
 \bar{V}[\rho , \hat{O}]_N \equiv 4 \,  \sum_{\tilde{\lambda}} p_{\tilde{\lambda}}  \, \Big(\sum_{x,y} \,  \bra{\tilde{\lambda}} \hat{o} (x) \hat{o} (y) \ket{\tilde{\lambda}}_{\mathrm{c}} \Big) \, ,
 \label{defC}
 \eeq
the average variance in a producible decomposition, as in Eq. \eqref{cpartmix31}, while the average variance 
in a generic decomposition $\{ p_{\lambda},\ket{\lambda}\} $ of $\rho$ (functionally defined as in Eq. \eqref{defC}\big) will be again denoted generically as $V[\rho , \hat{O}]_N$. The corresponding functionals involving the sums of the modula of the connected correlations will be denoted as $|V|[\rho , \hat{O}]_N$ and $|\bar{V}|[\rho , \hat{O}]_N$.

Starting from Eq. \eqref{cpartmix31} and exploiting the bound in Eq.  \eqref{FQineq}, we have that, see \cite{ren2021},
the inequality {\color{black} in Eq. \eqref{FQineq}}
holds also} for  $\bar{V}[\rho , \hat{O}]_N$ in Eq. \eqref{defC},
the average variance 
\beq 
\bar{V}[\rho , \hat{O}]_N  \leq 4 \,  k \,  \big[ c \, (N - h) + N \big]   \,  ,  
\label{finalboundmix}
\eeq
 in the presence of $(c,h)$-entanglement for the density matrix $\rho$ in Eq. \eqref{cpartmix31},} since $ \sum_{\tilde{\lambda}} \, \tilde{p}_{\lambda}  = 1$ \cite{notek,notasomma}.   
{\color{black} Actually, $(c,h)$-entanglement  is a sufficient but not necessary condition for the validity of Eq. \eqref{finalboundmix}.  Moreover, maximizing in $h$ the right-hand term in Eq. \eqref{finalboundmix}, the more common bound \cite{pezze2014}
\beq
 \bar{V}[\rho , \hat{O}]_N  \leq 4 \,  k \,  c \, N \, , 
 \label{boundsh}
 \eeq
 is obtained.
 Correspondingly, maximizing in $c$ the right-hand term in Eq. \eqref{finalboundmix}, $h$ is bound as $\bar{V}[\rho, \hat{O}]_N \leq 4 \, k \, \Big[(N-h+1)^2 + h - 1\Big]$.}
Eq. \eqref{finalboundmix} is obtained exploiting  the producibility of all the states $\ket{\tilde{\lambda}}$, with the same $c$ and $h$ (but not necessarily the same space-partition), so that the bound in Eq. \eqref{FQineq} holds for all of them. Instead, no orthonormality hypothesis of the $\ket{\tilde{\lambda}}$ set is required.
{\color{black} Moreover, note that the average variance for mixed states in a producible decomposition is related to the pure state variance Eq. \eqref{qidef} via
\beq
\bar{V}[\rho , \hat{O}]_N = \sum_{\tilde{\lambda}} \, p_{\tilde{\lambda}}  \, V[\ket{\tilde{\lambda}} , \hat{O}]_N \, .
\eeq
In this way, $\bar{V}[\rho , \hat{O}]_N$ also saturates the corresponding convexity inequality 
\beq
V[\rho , \hat{O}]_N \leq \sum_{\tilde{\lambda}} \, p_{\tilde{\lambda}}  \, V[\ket{\tilde{\lambda}} , \hat{O}]_N \, ,
\label{conv}
\eeq
valid in a generic decomposition \cite{pezze2014}.}

As it will be required later on, we introduce the quantity
\beq
| \bar{V}|[\rho , \hat{O}]_N \equiv 4 \,  \sum_{\tilde{\lambda}} p_{\tilde{\lambda}}  \, \Big(\sum_{x,y} \,  |\bra{\tilde{\lambda}} \hat{o} (x) \hat{o} (y) \ket{\tilde{\lambda}}_{\mathrm{c}}| \Big) \, ,
 \label{defCbar}
 \eeq
 (notice that $| \bar{V}|[\rho , \hat{O}]_N$ is not the modulus of
 $\bar{V}[\rho , \hat{O}]_N$). The same bounds in Eqs. \eqref{finalboundmix} and \eqref{boundsh} also hold, in the presence of $(c,h)$-entanglement for
 the density matrix $\rho$ in Eq. \eqref{cpartmix31}, for $| \bar{V}|[\rho , \hat{O}]_N$
 i.e., $\bar{V}[\rho , \hat{O}]_N \leq | \bar{V}|[\rho , \hat{O}]_N$, and 
 $| \bar{V}|[\rho , \hat{O}]_N \le 4 \,  k \, c \, N$, both for pure states and
 mixed states.
 The proof of the latter statement follows from the derivation
 presented in SM 2, where the bound
 $\bar{V}\le 4 \, k \, c \, N$ is shown for pure states. Then, the extension to mixed states is done as from Eq. \eqref{finalboundmix}
 to Eq. \eqref{defC}.

\subsection{Relation with the QFI}  
\label{relqfi}

For mixed states, the bounds in Eqs. \eqref{finalboundmix} and \eqref{boundsh} are also valid for the QFI \cite{pezze2014}.
This quantity is defined, in terms of the average variances in every decomposition $\{ p_{\lambda},\ket{\lambda}\}$ as follows 
\cite{yu2013,petz2013,toth2014,note1}:
\begin{equation}
F[\rho, \hat{O}]_N =  4 \,  \mathrm{Tr} \big[\rho \, \hat{O}^2] - 4 \sup_{\{ p_{\lambda},\ket{\lambda}\}} \sum_{\lambda} p_{\lambda}  \, \langle \lambda | \hat{O}  | \lambda \rangle^2   \, .
 \label{convroof_FQ}
\end{equation}
Notice that we are denoting by $\{\ket{\lambda}\}$ a generic decomposition, while
$\{\ket{\tilde{\lambda}}\}$ are the c-separable states entering Eq. (\ref{cpartmix31}).
  
\begin{figure}
\centering
  \includegraphics[scale=.25]{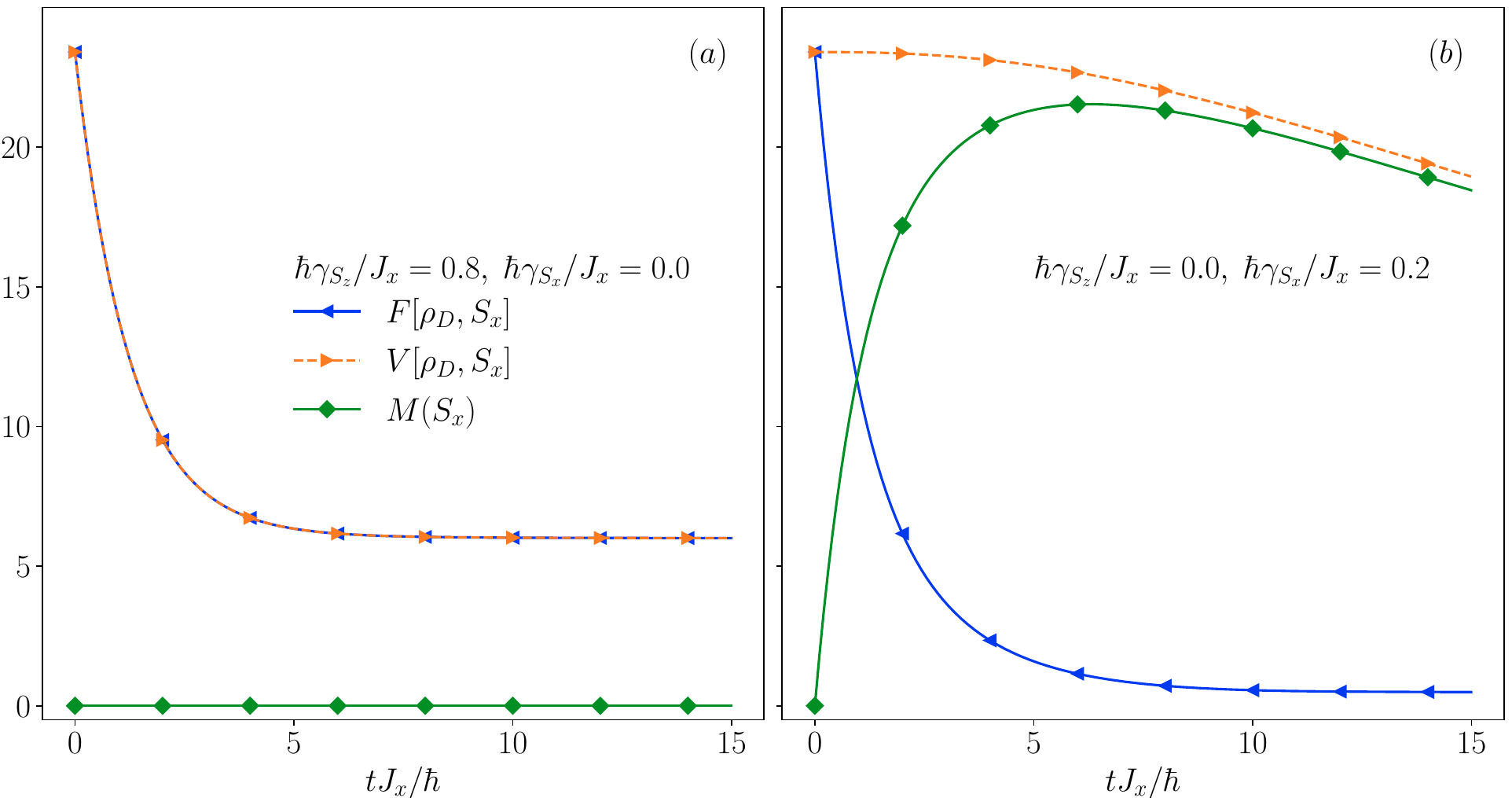}
\caption{Dissipative evolution of a XXZ chain of $N = 6$ site, using exact diagonalization, with a time-step of $J_{x}dt/\hbar = 0.01$, chosen to ensure numerical convergence. We begin the evolution from the ground state of $H$, with $S_z = 0$, in the gapless phase for $J_{z}/J_{x} = 0.8$, and study the evolution with (a) symmetry-respecting $\sigma^{(z)}_{i}$ dissipation, and (b) symmetry-breaking $\sigma^{(x)}_{i}$ dissipation. The observables for all panels are shown in the legend of (a). 
We also defined $M(S_x) \equiv V[\rho, S_x] - F [\rho, S_x]$, with both  $V[\rho, S_x]$ and $F [\rho, S_x]$ calculated in a diagonal decomposition.
}
\label{fig:Fig1}
\end{figure}

It turns out that  {\color{black} $\{\bar{V}[\rho, \hat{O}]_N , V[\rho, \hat{O}]_N \} \geq F[\rho, \hat{O}]_N$,} as formalized in the so-called "convex roof theorem" \cite{yu2013,petz2013}: for chosen $\rho$ and $\hat{O}$, the resulting QFI is the minimum average variance between all the possible decompositions 
$\{ p_{\lambda},\ket{\lambda}\} $ of $\rho$. 
The same property allows to demonstrate immediately that the bounds in Eqs. \eqref{finalboundmix} and \eqref{boundsh} also hold for the QFI, as it is well known in literature \cite{pezze2014}.
We also stress that the convex inequality in Eq. \eqref{conv} is fulfilled also by the QFI, that is saturated for pure states.\\
In a (spectral) decomposition $\ket{n}$, where $\rho$ is diagonal, the QFI, {\color{black} denoted in the following as $F[ \rho , \hat{O}]_N$,}  is written as \cite{pezze2014}
\beq
F[ \rho , \hat{O}]_N = 2 \sum_{x,y}  \sum_{n, m} 
\frac{\big(p_{n}  -  p_{m} \big)^2}{p_{n}  +  p_{m}}
 \langle n | \hat{o} (x) \ket{m} \bra{m} \hat{o} (y) | n \rangle \, .
\label{qfitherm0}
\eeq 
\indent  {\color{black} The described non-trivial relation between the QFI  $F[\rho , \hat{O}]_N$ and the average variances $V[\rho , \hat{O}]_N$ calculated in generic decompositions  $\{ p_{\lambda},\ket{\lambda}\} $} can be illustrated considering for instance a one-dimensional array of $N = 6$ 
spin-1/2, described  by the XXZ Heisenberg Hamiltonian in  a transverse magnetic field,
\begin{equation}
    H = -\sum_{i=1}^{N-1} \Big[ \frac{J_{x}}{2} \Big(\sigma_{i}^{(+)} \sigma_{i+1}^{(-)} + \sigma_{i}^{(-)} \sigma_{i+1}^{(+)} \Big)
+ J_{z} \, \sigma_{i}^{(z)} \sigma_{i+1}^{(z)}  \Big] - h_{x} S_{x} ~,
    \label{eq:H_XXZ}
\end{equation}
where $J_{x}$ and $J_{z}$ are exchange interactions, and $S_{\beta} = \sum_{i} \frac{\sigma^{(\beta)}_{i}}{2}$ ($\hbar \equiv 1$), $\beta = \{x, y, z \}$. For $h_x=0$, the total magnetization, $S_{z}=\sum_{i} \frac{\sigma_{i}^{(z)}}{2}$, is conserved, a fact exploited in \cite{daley2020} to compute the QFI efficiently. Due to this symmetry, the operator  $\hat{O}=S_x$ has $\bra{\lambda} \hat{O} \ket{\lambda} = 0,\, \forall \, \ket{\lambda}$ with  definite $S_{z}$. To analyze a mixed state scenario, we consider Markovian dissipation, according to a Gorini-Kossakowski-Sudarshan-Lindblad master equation \cite{lindblad1976, gorini1976, breuerpetruccione2002}, with local spin-flip and spin-dephasing noise described by jump operators $L_{m} = \{\sigma^{(x)}_{m}, \sigma^{(z)}_{m}\}$ respectively. Dissipation rates are denoted by $\gamma_{S_{x}}$ and $\gamma_{S_{z}}$. \\
\indent Fig.~\ref{fig:Fig1} shows the time evolution {\color{black} of the different functionals $F[ \rho , \hat{O}]_N$ and $V[ \rho , \hat{O}]_N$ (here both calculated in the diagonal decomposition),} starting from the ground state of $H$, with $S_z = 0$, in the presence of two forms of dissipation. 
In Fig.~\ref{fig:Fig1} (a), we include dissipation, as $L_{i} = \sigma^{(z)}_{i} $. As the $S_{z}$ symmetry is conserved throughout the evolution,  $\bra{\lambda} \sigma^{(x)}_{i} \ket{\nu} = 0$ at any time, {\color{black} then both $F[ \rho , \hat{O}]_N$ and $V[ \rho , \hat{O}]_N$ reduce to the first term in Eq. \eqref{convroof_FQ}, $4 \,  \mathrm{Tr} \big[\rho \, \hat{O}^2]$.}
In contrast, when the dissipation $L_{i} = \sigma^{(x)}_{i}$ does not preserve the magnetization, as shown in Fig.~\ref{fig:Fig1} (b), {\color{black} then the QFI quickly differs from  $V[\rho , \hat{O}]_N$, the last quantity being higher in value, as expected from the convex roof theorem.}

\section{Discussion of conditions to estimate multipartite entanglement by correlation functions}
\subsection{The general problem}
 The producible decompositions $\ket{\tilde{\lambda}}$ and $p_{\tilde{\lambda}}$, where $\bar{V}[\rho , \hat{O}]_N$ is defined as in \eqref{defC}, are not generally known \textit{a priori}. Thus, in order to be a useful witness of ME,
$\bar{V}[\rho , \hat{O}]_N$ is required to be calculable without the knowledge of $\ket{\tilde{\lambda}}$. To focus on this central problem,
it is useful to decompose $\bar{V}[\rho , \hat{O}]_N$ into two terms: $\bar{V}[\rho , \hat{O}]_N \equiv \bar{V}_1[\rho , \hat{O}]_N - \bar{V}_2 [\rho , \hat{O}]_N$. 
 The first term, 
$ \bar{V}_1[\rho , \hat{O}]_N \equiv 4 \sum_{x,y} \sum_{\tilde{\lambda}} p_{\tilde{\lambda}} \, \bra{\tilde{\lambda}} \hat{o} (x) \hat{o} (y) \ket{\tilde{\lambda}}$, can be recast as
\beq
 \bar{V}_1[\rho , \hat{O}]_N = 4 \, \sum_{x,y} \mathrm{Tr} \big[ \rho \, \hat{o} (x) \hat{o} (y) \big] = 4 \,  \mathrm{Tr} \big[\rho \, \hat{O}^2] 
\label{F1}
\eeq
{\color{black} (the first term in Eq. \eqref{convroof_FQ}\big),}  where $\rho$ and $\hat{o} (x)$ are meant to be expressed in a generic basis $\ket{\alpha}$ of the Hilbert space, possibly orthonormal, as $\ket{\tilde{\lambda}} = \sum_{\alpha} c_{\tilde{\lambda} \, \alpha} \ket{\alpha}$.
This expression does not depend explicitly on $\ket{\tilde{\lambda}}$, thus it is covariant,
  and invariant in value, under changes of  decomposition. This transformation is realized via a unitary operator $U$, acting as {\color{black} (see  SM 3)}: 
\beq
\sqrt{p_{\tilde{\lambda}}} \, \ket{\tilde{\lambda}} = U_{\tilde{\lambda} \, \lambda} \, \sqrt{p_{\lambda}} \,  \ket{\lambda} \, .
\label{cambiodec}
\eeq 
In contrast, the second term 
\beq
 \bar{V}_2[\rho , \hat{O}]_N \equiv  4 \, \sum_{x,y} \sum_{\tilde{\lambda}} p_{\tilde{\lambda}} \, \langle \tilde{\lambda} | \hat{o} (x) | \tilde{\lambda} \rangle  \langle \tilde{\lambda}  | \hat{o} (y) | \tilde{\lambda} \rangle \geq 0
 \label{V2}
 \eeq 
  is neither invariant nor covariant under the transformation $U$ in general,  
hence to evaluate it requires the explicit knowledge of $\ket{\tilde{\lambda}}$. In turn, the non-covariance of $ \bar{V}_2[\rho , \hat{O}]_N$ makes it difficult to calculate $ \bar{V}[\rho , \hat{O}]_N$ in a generic decomposition, and to use it
as an efficient ME estimator. 
The same problem occurs in calculating the QFI, via  $\big[ \sup_{\{ p_{\lambda},\ket{\lambda}\}} \sum_{\lambda} p_{\lambda}  \, \langle \lambda | \hat{O}  | \lambda \rangle^2 \big]$ in Eq. \eqref{convroof_FQ}. This is the main reason why the cumbersome expression in Eq. \eqref{qfitherm0} must be used, in general.

\subsection{{\color{black} Two conditions for the calculability} of Eq. \eqref{defC}}
\label{cond}
In the present subsection, we discuss {\color{black} two situations when} the quantity
\beq
\sum_{\lambda} p_{\lambda}  \, \langle \lambda | \hat{O}  | \lambda \rangle^2 = \sum_{\lambda} p_{\lambda}  \sum_{x,y} \, \bra{\lambda} \hat{o} (x) \ket{\lambda} \bra{\lambda} \hat{o} (y) \ket{\lambda} 
\label{secondot}
\eeq
in Eq. \eqref{convroof_FQ} can be calculated, in a generic decomposition $\ket{\lambda}$, in spite of the difficulties mentioned above.

-- Let first consider $\hat{O}$ as a semipositive or seminegative defined operator in the considered Hilbert space.
We recall that an operator $\hat{o}$ is called  positive definite on a given Hilbert state if,
for any vector $\ket{v}$ on this space, $\bra{v} \hat{o} | v \rangle > 0$. 
Similarly, it is called  semi-positive definite if $\bra{v} \hat{o} | v \rangle \geq 0$. 
We mention for instance  collective spin operators $\big(S^{(x,y,z)} \big)^2 = \sum_i \big(s_i^{(x,y,z)}\big)^2$ for $s > \frac{1}{2}$ or $\tilde{S}^{(x,y,z)} = \sum_i  \big(\pm s \, {\bf I} + s_i^{(x,y,z)} \big) $ for $s \geq \frac{1}{2}$ (notice that  $F[\rho , a \, {\bf I} +\hat{O}]_N =  F[\rho , \hat{O}]_N$).  In this condition,
if $\langle \lambda | \hat{O}  | \lambda \rangle = 0$,  for any state $\ket{\lambda}$ in a chosen decomposition, then the  quantity in Eq. \eqref{secondot} vanishes in any other decomposition, since also $\sum_{\lambda} \,  \langle \lambda | \hat{O}  | \lambda \rangle = \mathrm{Tr} \big[ \rho \,  \hat{O}  \big] = 0$ and $\mathrm{Tr} \big[ \rho \,  \hat{O}  \big]$ is  invariant under changes of decompositions.\\
 
-- Second, let us consider states with the property that
\beq
\bra{\lambda} \hat{o} (x) \ket{\lambda} \equiv o_{\lambda}
\label{indep}
\eeq
is independent of $x$, for the chosen local operator $\hat{o} (x)$ and for every $\ket{\lambda}$ of a certain decomposition.
{\color{black} Translationally invariant states can be included, see more details below.}
In this condition, the quantity in Eq. \eqref{secondot} is evaluated as follows.
{\color{black} Defining} $ \mathrm{Tr}[\rho \, \hat{o}(x) \hat{o} (y)] \equiv c(x,y)$ (appearing in $\bar{V}_1[\rho , \hat{O}]_N$, Eq. \eqref{F1}, and always depending on $x$ and $y$),
if 
\beq
\sum_{\lambda} p_{\lambda}  \, \langle \lambda | \hat{O}  | \lambda \rangle^2 = N^2 \, \mathrm{lim}_{|x-y| \to \infty} \, c(x,y) \equiv  N^2 \, c_{\infty} \, ,
\label{secondot2}
\eeq
{\color{black} then it is possible to recast the average variance $V[\rho ,\hat{O}]_N$, in the present decomposition $\{ p_{\lambda},\ket{\lambda}\} $, as}
\be
V[\rho , \hat{O}]_N =4 \,  \Big[ \sum_{x,y}  c(x,y) - N^2 \,  c_{\infty} \Big] \, .
\label{eq:decomposition}
\ee 
The conditions for the validity of Eq. \eqref{secondot2} will be discussed in the next subsection.

Although evaluated in a specific decomposition, this functional
will be used to determine conditions under which it is possible
to bound multipartite entanglement for the mixed state described by $\rho$.
This
result will be shown in the next Section, as well as the relation of the same functional
with $\bar{V}[\rho , \hat{O}]_N$ in Eq. \eqref{defC}  and with the QFI, $F[\rho, \hat{O}]_N$.

\subsection{Derivation 
  of Eq. (19)}

In this Section, we derive and motivate Eq. \eqref{eq:decomposition}. In order to perform this task,
we have to discuss first the conditions for the validity of Eq. \eqref{secondot2}.
 For this purpose, beyond Eq. \eqref{indep},
the second property that we assume at first is:
\beq
\bra{\lambda} \hat{o} (x) \hat{o} (y) \ket{\lambda} \to \bra{\lambda} \hat{o} (x) \ket{\lambda} \bra{\lambda} \hat{o} (y) \ket{\lambda}   \, ,
\label{cdec}
\eeq
if $|x-y| \to \infty$~\cite{hastings2004,sims2006,weinberg1}. However, this property will be relaxed in the following subsections, {\color{black} since eventually it will not be strictly
required} to justify Eq. \eqref{eq:decomposition} as an entanglement witness.

Notice that the simultaneous validity of Eqs. \eqref{indep} and  \eqref{cdec} requires strictly that
$\bra{\lambda} \hat{o} (x) \hat{o} (y) \ket{\lambda}$ does not depend on the unit vector of $x-y$, nor on $x$ and $y$ themselves, at least if $|x-y| \to \infty$. 
Sufficient, but not necessary, conditions for this scenario are rotational invariance and again translational invariance.

We also stress that, in order to guarantee that the one-point correlations are not point-dependent, here  translational invariance  must be understood for every translation $a_i$, connecting two sites of the lattice and for every state $\ket{\lambda}$. For regular lattices, these translations decompose into sums of translations inside a unit cell and primary lattice vectors, see e. g. \cite{gp}. {\color{black} Finally, if the condition in Eq. \eqref{cdec} holds for every local operator  $\hat{o} (x)$, this is called "cluster decomposition" in the literature.}

Importantly, Eq. \eqref{cdec} does not imply at all that $\mathrm{Tr} \big[ \rho \, \hat{o} (x) \hat{o} (y) \big]$ tends, even for large
space separations, to the product $\mathrm{Tr} \big[ \rho \, \hat{o} (x) ) \big]  \mathrm{Tr} \big[ \rho \,  \hat{o} (y) \big]$: this can be seen also as an effect
of the classical weights $p_{\lambda}$, not encoding entanglement. Moreover, the fact that $o_{\lambda}$ is not $x$-dependent implies the same fact for $\mathrm{Tr} \big[ \rho \, \hat{o} (x) ) \big] = \sum_{\lambda} p_{\lambda} o_{\lambda}$. The opposite implication does not hold in general, requiring further assumptions, as translational invariance. 

We also comment that the latter property sets a tight limitation to the degree of producibility $c$, even for a single pure state $\ket{\psi}$: indeed, 
consider two adjacent points $x$ and $y$: as argued in  Section \ref{bme}, if they belong to the same entangled subset $\ket{\psi_i}$, then $\bra{\lambda} \hat{o} (x) \hat{o} (y) \ket{\lambda}_{\mathrm{c}}$ is nonzero in general, otherwise the same quantity is forced to vanish. Translational invariance,
for every translation $a_i$, connecting two sites of the lattice, implies
\beq
\bra{\lambda} \hat{o} (x + a_i) \hat{o} (y + a_i) \ket{\lambda}_{\mathrm{c}} = \bra{\lambda} \hat{o} (x) \hat{o} (y) \ket{\lambda}_{\mathrm{c}} \, .
\eeq
From the discussion in Section \ref{bme}, the latter equality implies immediately that $c = 1$ or $c = N$.
More involved scenarios would be possible instead if translational invariance was allowed only for a subset of the entire set of lattice-translation $\{a_i\}$.

Under the two conditions in Eqs. \eqref{indep} and \eqref{cdec}, the quantity in Eq. \eqref{secondot} can be evaluated straightforwardly, leading to Eqs. \eqref{secondot2}  and \eqref{eq:decomposition} (see more details in SM 5).
We stress that the same derivation exploits that all the states of the decomposition  $\ket{\lambda}$ fulfill Eq. \eqref{cdec}. 
However, the same property ceases to hold in general, as the decomposition is changed along Eq. \eqref{cambiodec}. Consequently,  one can verify that Eq.
\eqref{eq:decomposition} is not equal to  $\bar{V}[\rho , \hat{O}]_N$ nor to $F[\rho, \hat{O}]_N$, in general. Nevertheless, the same functional
\eqref{eq:decomposition} will reveal still useful to bound multipartite entanglement in any decomposition.

\subsection{Comments on the space independence of the one-point correlations}

Here, having in mind the point {\it ii)} of Section \ref{cond}, we  discuss under which conditions the independence of the one-point correlations on the point itself can
hold in every decomposition.

 At first, we consider, in a given decomposition $\ket{\lambda}$, the matrix elements
$\bra{\lambda} \hat{o} (x) \ket{\lambda}$ and  $\bra{\lambda} \hat{o} (y) \ket{\lambda}$, $y \neq x$, and we assume the stronger condition of translational invariance.
{\color{black} Note that we do not a priori require that these conditions hold in every decomposition. We can interpolate between the two matrix elements above,} writing:
\beq
\bra{\lambda} \hat{o} (y) \ket{\lambda} = \bra{\lambda}  \hat{T}_{y-x} \, \hat{o} (x) \, \hat{T}^{-1}_{y-x} \ket{\lambda} \, ,
\eeq
where $\hat{T}_{y-x} = e^{i \, \hat{P} \, (y-x)}$ is a unitary translation operator, generated by the momentum operator $\hat{P}$. Here $y-x$ is a multiple number of the lattice steps, possibly even in different primary directions for regular lattices.
If $\ket{\lambda}$ in translationally invariant, then $\hat{T}_{y-x} \, \ket{\lambda} = e^{i \, k \, (y-x)} \ket{\lambda}$, $k$ being the momentum quantum number.
The space-dependent phases cancel out in the matrix element  $\bra{\lambda} \hat{o} (y) \ket{\lambda}$, so that, in the end:
\beq
\bra{\lambda} \hat{o} (y) \ket{\lambda} = \bra{\lambda}  \hat{o} (x) \ket{\lambda} \, , 
\eeq
as expected. Notice that, critically, the same cancellation does not occur in general for the off-diagonal
matrix elements $\bra{\lambda^{\prime}} \hat{o} (y) \ket{\lambda}$, $\ket{\lambda^{\prime}} \neq \ket{\lambda}$.

Consider now a second decomposition $\ket{\eta}$: this is obtained from $\ket{\lambda}$ as in Eq. \eqref{cambiodec}, so that:
\beq
\bra{\eta} \hat{o} (y) \ket{\eta} = \sum_{\lambda \, , \, \lambda^{\prime}} \frac{\sqrt{p_{\lambda} p_{\lambda^{\prime}}}}{p_{\eta}} \, [U^*]_{\lambda^{\prime} \, \eta} U_{\eta \, \lambda} \, \bra{\lambda^{\prime}}  \hat{o} (y) \ket{\lambda} \, .
\label{cambiodec2}
\eeq
Importantly, the involved unitary matrix $U_{\eta \, \lambda}$ is not space dependent, as well as the numerical factors $\sqrt{p_\eta}$ and $\sqrt{p_\lambda}$. 
Therefore, the only difference between the elements $\bra{\eta} \hat{o} (y) \ket{\eta}$ and $\bra{\eta} \hat{o} (x) \ket{\eta}$ can come from the phases $e^{i \, k \, (y-x)}$, $e^{i \, k^{\prime} \, (y-x)}$,
not canceling out in the off-diagonal matrix elements $\bra{\lambda^{\prime}}  \hat{o} (y) \ket{\lambda}$, $\ket{\lambda^{\prime}} \neq \ket{\lambda}$.

We conclude that, if the states of the decomposition $\ket{\lambda}$ are such that $\bra{\lambda} \hat{o} (x) \ket{\lambda}$ does not depend on $x$,
the same property holds for another related decomposition $\ket{\eta}$ (at least) if $\bra{\lambda^{\prime}} \hat{o} (x) \ket{\lambda} = 0$ when  $\ket{\lambda^{\prime}} \neq \ket{\lambda}$ and $\forall \, x$.
A required, but not sufficient, condition, is clearly that, in the same condition,  also  $\bra{\lambda^{\prime}} \hat{O} \ket{\lambda} = 0$.
Due to Eq. \eqref{cambiodec2}, the same conclusion is reached in the more general case when the independence of $\bra{\lambda} \hat{o} (x) \ket{\lambda}$ is realized
without the strongest assumption of translational invariance {\color{black} (still provided the vanishing of the off-diagonal matrix elements, assumed $x$-dependent).} 

Various examples are plausible  with the feature described above.
Consider for instance a density matrix with  an orthogonal decomposition (as a spectral one) $\ket{\lambda}$,
such that its elements are eigenstates of $\hat{O} = S_{z} = \sum_{i} \sigma_{i}^{(z)}$ ($i$ labelling the sites, now), as in the example of Section \ref{relqfi}, {\color{black} with periodic boundary conditions.} However, now all these states are characterized by different
eigenvalues of $\hat{O}$. Since clearly $\big[ S_{z} , \sigma_{i}^{(z)} \big] = 0$, then $\sigma_{i}^{(z)} \ket{\lambda} = \sum_{\lambda^{''}} a_{\lambda , \lambda^{''}} \, \ket{\lambda^{''}} $,
where $\ket{\lambda^{''}}$  are eigenstates of $S_{z}$ with the same eigenvalue of $\ket{\lambda}$. In the subspace spanned by the set  $\ket{\lambda^{''}}$, the same states $\ket{\lambda^{''}}$ can be always assumed to be arranged orthogonal to $\ket{\lambda^{\prime}} \neq \ket{\lambda}$. Therefore $\bra{\lambda^{\prime}} \sigma_{i}^{(z)} \ket{\lambda} = 0$.

{\color{black}
\subsection{Further comments on the property in Eq. \eqref{cdec}}
\label{clustersec}

 In the following, we analyze some consequences of Eq.  \eqref{cdec}, relevant for our purposes.
 In particular,} the same equation immediately implies, and is implied by, that, for a certain pure state $\ket{\psi}$, the scaling of $V[\ket{\psi} , \hat{O}]_N$ with $N$ is below the Heisenberg one. 
Similarly, if the average variance $V[\rho ,\hat{O}]_N$ in a certain decomposition $\{ p_{\lambda},\ket{\lambda}\} $
 can be written in the invariant form \eqref{eq:decomposition} (then it is possible to set $ \sum_{\lambda} p_{\lambda} \, o_{\lambda}^2 = c_{\infty}$), its scaling with $N$ is below the Heisenberg one, and vice versa.
 
 These facts hold because, if 
\beq
|\langle \psi | \hat{o}(x) \hat{o}(y)| \psi \rangle_{\mathrm{c}}| \sim |x-y|^{- 2\alpha} ~~ \mathrm{when} ~~ |x-y| \to \infty \, ,
\label{scalcor}
\eeq
then  $|V|[\ket{\psi}, \hat{O}]_N$, scales with $N \to \infty $ as \cite{hauke2016,pezze2017,lepori2022}:
\beq
|V|[\ket{\psi}, \hat{O}]_N \sim N^{2 - 2 \frac{\alpha}{d}}  \, .
\label{qfiscaling}
\eeq
Notice that, as required by the simultaneous validity of Eqs. \eqref{indep} and  \eqref{cdec} (see Section \ref{clustersec}), in Eq. \eqref{scalcor} we still assumed that
$\bra{\lambda} \hat{o} (x) \hat{o} (y) \ket{\lambda}$ does not depend on the unit vector of $x-y$, nor on $x$ and $y$ themselves, at least if $|x-y| \to \infty$.
Conversely, Eq. \eqref{scalcor} implies  the 
scaling $|\langle \psi | \hat{o}(x) \hat{o}(y) | \psi \rangle_{\mathrm{c}}| \sim |x-y|^{-2\alpha}$. Moreover, Eq. \eqref{qfiscaling} completes the previously known result (see e.g. \cite{hauke2016,pezze2017}) that $|V|[\ket{\psi}, \hat{O}]$
increases extensively with $N$ if  $|\langle \psi | \hat{o}(x) \hat{o}(y) | \psi \rangle_{\mathrm{c}}|$ decays exponentially with $|x-y|$, as for short-range gapped systems. 

It is immediate to check, by direct analogy, that the same proof of  Eq. \eqref{qfiscaling} \cite{lepori2022} holds for  
the link between $|V|[\rho ,\hat{O}]_N$ and the argument
\beq
\sum_{\lambda} p_{\lambda}  \,   |\bra{\lambda} \hat{o} (x) \hat{o} (y) \ket{\lambda}_{\mathrm{c}} | 
\eeq
of the sum in $x$ and $y$ in its definition, as in Eq. \eqref{defCbar}.

From Eq. \eqref{qfiscaling}, it results that  the "clustering" property in Eq. \eqref{cdec} ($\alpha \geq 0$ in Eq. \eqref{scalcor}) is  generally required,  but not sufficient (since another operator $\hat{o}(x)$ could evade it),  to have $c < N$, $c$ bounded as in Eq.~\eqref{FQineq}.
Instead, its violation, 
$\alpha \leq 0$, 
 implies $c = N$.
{\color{black} Actually, even when the point-independence holds of the one-point correlations  in Eq. \eqref{indep}, the described difficulty to evaluate $ \bar{V}_2[\rho , \hat{O}]_N$ in Eq. \eqref{V2} (and $c_{\infty}$ in a generic decomposition $\{ p_{\lambda},\ket{\lambda}\} $) is directly related with the general violation of Eq. \eqref{cdec}.}

{\color{black} Finally, we notice that a lower bound for the scaling of $c$ can be obtained further from the failure of Eq. \eqref{cdec} and without exploiting the scaling of $|V|[\ket{\psi}, \hat{O}]_N$: $c \sim N^{\gamma}$, $\gamma \geq {(d-1)}/{d}$, see SM 4. This bound reflects the entanglement between a bulk point and the boundary of the lattice implied by the failure of Eq. \eqref{cdec}. Moreover, it demonstrates $c \sim N$ in the mean-field limit $d \to \infty$. }

\section{Use of the functional in Eq. (\ref{eq:decomposition}) to bound multipartite entanglement}

We are not able to provide constraints on a density matrix $\rho$, sufficient to assure that the property in Eq. \eqref{cdec} holds in any decomposition. However, this task 
will turn out not to be strictly required for our purposes.

Indeed, in the present
Section, we show how the functional in Eq. \eqref{eq:decomposition}
can be exploited to bound multipartite entanglement, provided that the property in Eq. \eqref{indep} holds for each decomposition - even without invoking the cluster decompistion property (\ref{cdec}).

 Two main situations can occur:

\indent -- $c < N$: in this case, the property in Eq. \eqref{cdec} holds for the decomposition $\{ p_{\tilde{\lambda}},\ket{\tilde{\lambda}}\} $ in Eq. \eqref{cpartmix31}, from the discussion of the previous subsection. Therefore, the average variance  
$\bar{V}[\rho , \hat{O}]_N$ can be expressed in the form of Eq. \eqref{eq:decomposition}. The same expression is covariant and invariant under changes of decomposition
(therefore, it can be equivalently evaluated in  $\{ p_{\tilde{\lambda}},\ket{\tilde{\lambda}}\} $ or in any other decomposition),
and it is also equal to the QFI in the present conditions:
\beq
\bar{V}[\rho , \hat{O}]_N = V[\rho , \hat{O}]_N =F[\rho , \hat{O}]_N \, .
\label{consclust}
\eeq
Indeed, referring to Eq. \eqref{convroof_FQ}: 
\beq
\sup_{\{ p_{\lambda},\ket{\lambda}\}} \sum_{\lambda} p_{\lambda}  \, \langle \lambda | \hat{O}  | \lambda \rangle^2  = \inf_{\{ p_{\lambda},\ket{\lambda}\}} \sum_{\lambda} p_{\lambda}  \, \langle \lambda | \hat{O}  | \lambda \rangle^2  = c_{\infty} \, .
\eeq
Therefore, Eq. \eqref{eq:decomposition} can be exploited to bound, via the bounds in the subsections \ref{bme},  the actual value of $c$. Clearly, this value must be lower than $N$, as implied also by the construction of Eq. \eqref{eq:decomposition}). 

\indent -- $c = N$: in this case, if the property in Eq. \eqref{cdec} holds, everything works as in the previous case, apart from the fact that now the estimated value of $c$ by
the functional $V[\rho , \hat{O}]_N$ in Eq. \eqref{eq:decomposition} can saturate to $N$. 

Instead, if Eq. \eqref{cdec} does not hold, the same functional (not sharing any clear general relation with  $F[\rho , \hat{O}]_N$, at the best of our understanding), still calculated in a chosen working decomposition, can yield a lower value $c<N$ (again by construction of Eq. \eqref{eq:decomposition}). Therefore, a lower bound for $c$ is still established.  \\

\section{Conclusions}  
We discussed {\color{black} some physical and mathematical conditions that make it possible} to bound multipartite entanglement for
mixed quantum states, via 
one- and two-point correlation functions.

Our analysis holds for discrete systems, 
but it could be extended -- although not straightforwardly -- to continuous systems \cite{gessner2016,gessner2017_2}.  Further investigation
is required to identify other physically interesting cases, where our approach can be useful.\\

\section*{Acknowledgments} 
The authors are pleased to thank Michele Burrello and Luca Pezz\`e for many fruitful discussions. \\
 L. L.  and D. G. acknowledge financial support by the PRIN project number 20177SL7HC, financed by the Italian Ministry of education and research. \\
 L. L.  and A. S. acknowledge financial support by the Qombs Project [FET Flagship on Quantum Technologies grant n. 820419]. \\
 L. L. acknowledge financial support by a project funded under the National Recovery and Resilience Plan (NRRP), Mission 4 Component 2 Investment 1.3 - Call for tender No. 341 of 15/03/2022 of Italian Ministry of University and Research funded by the European Union NextGenerationEU, award number PE0000023, Concession Decree No. 1564 of 11/10/2022 adopted by the Italian Ministry of University and Research, CUP D93C22000940001, Project title ”National Quantum Science and Technology Institute” (NQSTI), spoke.\\
 This study was also carried out within the National Centre on HPC, Big Data and Quantum Computing - SPOKE 10 (Quantum Computing) and received funding from the European Union Next-GenerationEU - National Recovery and Resilience Plan (NRRP) – MISSION 4 COMPONENT 2, INVESTMENT N. 1.4 – CUP N. I53C22000690001.\\
 M. L. Chiofalo acknowledges support from the MIT-UNIPI program. \\
 Work at the University of Strathclyde was supported by the EPSRC Programme Grant DesOEQ (EP/P009565/1), and by the European Union’s Horizon 2020 research and innovation program, under grant agreement No.~817482 PASQuanS.

\newpage

\onecolumngrid

\appendix

\section*{SUPPLEMENTARY MATERIAL}

\section*{SM 1: QFI and ME for mixed states}

In this Section, we recall some elements of the general theory of the quantum Fisher information (QFI) \cite{pezze2014},
a central tool in estimation theory \cite{pezze2014}. 

This is defined via the fidelity between a pure quantum state $\ket{\psi(\theta)}$, depending on a parameter $\theta$, and its infinitesimal variation $\ket{\psi(\theta + \mathrm{d} \theta)}$:
$\mathcal{F} (\theta , \mathrm{d} \theta) = |\bra{\psi(\theta)} \psi(\theta +  \mathrm{d} \theta) \rangle |^2  \approx 1 - F(\theta) {(\mathrm{d}\theta)^2}/{8}$.
Among the possible transformations relating $\ket{\psi(\theta)}$
and $\ket{\psi(\theta + \mathrm{d} \theta)}$ on discrete systems, like  numerable sets of qubits or lattices (in the following labelled by $x$ and $y$), a relevant choice is the unitary evolution generated by a collective Hermitian operator $\hat{O} = \sum_x \hat{o}(x)$: 
$\ket{\psi(\theta + \mathrm{d} \theta)} = e^{i  \, \mathrm{d} \theta \, \hat{0}} \ket{\psi(\theta)}$.

For $d$-dimensional $N$-component discrete systems, the QFI  for pure states equals the variance $V[\ket{\psi}, \hat{O}]_N$ in Eq. (3)
of the main text, and it can be cast in terms of two-point {\it connected} correlations (the $\theta$ dependence being neglected)~\cite{noteconv}
\be
\langle \psi | \hat{o} (x) \hat{o} (y) | \psi \rangle_{\mathrm{c}} \equiv  \langle \psi | \hat{o} (x) \hat{o} (y) | \psi \rangle - \langle \psi | \hat{o} (x) | \psi \rangle \langle \psi | \hat{o} (y) | \psi \rangle \, ,
\ee
as
\beq
F[\ket{\psi}, \hat{O}]_N = V[\ket{\psi}, \hat{O}]_N = 4\sum_{x,y}\langle \psi | \hat{o} (x) \hat{o} (y) | \psi \rangle_{\mathrm{c}} \, . \label{qidef}
\eeq

For two mixed states, described by density matrices $\rho$ and $\sigma$, the fidelity, similar to that for pure states, is defined as \cite{uhlmann1976,pezze2014}:
\beq
\mathcal{F}(\rho, \sigma) = \Big( \mathrm{Tr} \,  \sqrt{\sqrt{\rho} \, \sigma \sqrt{\rho}}  \Big)^2 \, \, .
\eeq
The last expression is considered applied to two infinitesimally-separated density matrices  
\beq
\rho (\theta) = \sum_{\lambda} p_{\lambda} \,  \rho_{\lambda} (\theta) \, ,  \quad \rho_{\lambda} (\theta) = \ket{\lambda(\theta)} \bra{\lambda(\theta)}   \, ,
\label{densmix}
\eeq
 and $\rho (\theta + \mathrm{d} \theta) =  e^{- i  \, \mathrm{d} \theta \, \hat{O}}  \rho(\theta) \,  e^{i  \, \mathrm{d} \theta \, \hat{O}} $, $\hat{O}$ labelling a generic Hermitian operator. There,
$0 \leq p_{\lambda} \leq 1$, and the  basis  $\ket{\lambda}$ is chosen orthonormal,  then $\rho(\theta) \equiv \rho_D$ is in its diagonal form. 
The resulting QFI, $- 2 \, \frac{\mathrm{d}^2 \mathcal{F}(\theta, \mathrm{d} \theta)}{\mathrm{d} \theta^2} $, reads:
\beq
F[\rho_D , \hat{O}]_N = 2 \, \sum_{x,y}  \sum_{\lambda, \nu} \, 
\frac{\big(p_{\lambda}  -  p_{\nu} \big)^2}{p_{\lambda}  +  p_{\nu}}  \,
 \langle \lambda | \hat{o} (x) \ket{\nu} \bra{\nu} \hat{o} (y) | \lambda \rangle   \, ,
\label{qfitherm0bis}
\eeq
that for a pure state clearly reduces to the variance $V[\ket{\psi}, \hat{O}]_N$  in Eq. (3)
of the main text, since in that case $p_{\lambda} = 1$ just for a certain state $\ket{\lambda}$, vanishing for the other ones.
The QFI in Eq. \eqref{qfitherm0bis} fulfills the inequality \cite{pezze2014}
\beq
F[\rho_D , \hat{O}]_N  \leq \sum_{\lambda}  \, p_{\lambda}  \, F[\ket{\lambda}, \hat{O}]_N	\, ,  
\label{ineqFs}
\eeq
$p_{\lambda} \geq 0$ and the bound being saturated by pure states. This convexity inequality is strictly required by an entanglement witness, since physically  this property reflects the fact that mixing quantum states cannot increase the entanglement content, as well as the related achievable estimation sensitivity \cite{pezze2014}. 

The QFI cannot be expressed entirely in terms of two-point connected correlation functions, indeed
the following relation holds \cite{gabbrielli2018,wang2014}
\beq
F[\rho_D , \hat{O}]_N = 4 \,  \sum_{x,y} \Bigg( \sum_{\lambda} p_{\lambda} \bra{\lambda} \hat{o} (x) \hat{o} (y) \ket{\lambda}_{\mathrm{c}} -  2  \sum_{\lambda, \nu \neq \lambda}  \, \frac{p_{\lambda} \,   p_{\nu} }{p_{\lambda}  +  p_{\nu}}   \, \langle \lambda | \hat{o}(x) \ket{\nu} \bra{\nu} \hat{o}(y)| \lambda \rangle \Bigg)  \, .
\label{qieqbis}
\eeq

Part of the recognized importance of the QFI \eqref{qidef}  is that it witnesses ME, both for pure and mixed states 
\cite{PezzePRL2009, HyllusPRA2012, TothPRA2012}. 
Indeed, the violation of the inequality 
\be 
F [\rho_D , \hat{O}]_N \leq 4 k \Big[ c (N-h) + N \Big]  \, ,
\label{FQineqsupp}
\ee
where   $ \frac{N}{c}   \leq h \leq N - c +1$ is the number of disentangled subsets, signals at least $(c+1)-$partite entanglement between the $N$ components of the considered system, with $1\leq c \leq N$ \cite{ren2021}. 
In \eqref{FQineqsupp}, $k = \frac{(m - n)^2}{4}$, if  $\hat{o} (x)$ is bounded, with eigenvalues $n \leq  \lambda  \leq m < \infty$.
Actually, $c$ can also diverge with $N$ $c \sim N^{l}$. 
The ultimate limit  $V[\vert \psi \rangle, \hat{O}]_N \, = N^2$ is called the Heisenberg limit and $\ket{\psi}$ is then said to host {\it genuine} ME.
Eq. \eqref{FQineqsupp} generates other relevant bounds \cite{ren2021}: for instance, it can be increased choosing $h =  \frac{N}{c} $,
which yields the more common bound ~\cite{PezzePRL2009,pezze2014}
\beq 
F[\rho_D , \hat{O}]_N \leq 4 \, k \, c \, N \, ,
\eeq
useful if $h$ is unknown. 
A direct derivation of this bound  for pure states will be given in SM 2.

{\color{black}
\section*{SM 2: derivation of the bound in Eq. (8) of the main text for pure states}
\label{proofbound}

In this Section, we discuss the bounds for pure states in 
the main text, presenting a treatment
different from the original one in \cite{PezzePRL2009}.
In particular, the present discussion is valid for the quantum Fisher
information (QFI) $F[\ket{\psi}, \hat{O}]_{N}$
only on pure states, and proceeds by the direct analysis of
the two-point connected correlation function. We recall that, for a pure state $\ket{\psi}$,
\beq
F[\ket{\psi}, \hat{O}]_{N} = V[\ket{\psi}, \hat{O}]_{N} = \bar{V}[\ket{\psi}, \hat{O}]_{N} \, ,
\eeq
$V[\ket{\psi}, \hat{O}]_{N}$ and  $\bar{V}[\ket{\psi}, \hat{O}]_{N}$ being defined as in the main text.

We start from the verified \cite{shi2003,shi2004} assumption that,
  if entanglement does not hold between two different sets, all the  connected correlation between them (the discrete components, e.g. the sites) must vanish, and one can rewrite the quantum Fisher information on a generic pure state $\ket{\psi}$ as follows:
\beq
F[\ket{\psi}, \hat{O}]_{N} =4 \sum_{D_a} \sum_{i,j \in D_a}  \langle \psi | \hat{o}_i \hat{o}_j| \psi \rangle_{\mathrm{c}} \, ,
\label{qidef5}
\eeq
where $D_a$ are the $h$ unentangled subsets of $\ket{\psi}$.   
If the decomposition in Eq. \eqref{qidef5} holds, exploiting the triangular inequality, Eq. \eqref{qidef5} can be bounded as follows:
\beq
F[\ket{\psi}, \hat{O}]_{N} \leq  4 \sum_{D_a} |\sum_{i,j \in D_a}  \langle \psi | \hat{o}_i\hat{o}_j| \psi \rangle_{\mathrm{c}} | \leq  4 \sum_{D_a} \sum_{i,j \in D_a} | \langle \psi | \hat{o}_i\hat{o}_j| \psi \rangle_{\mathrm{c}} | \leq  4 \sum_{D_a}  n_a^2  \, \,  \mathrm{max}_{i, j \in D_a} |\langle \psi | \hat{o}_i\hat{o}_j| \psi \rangle_{\mathrm{c}}  | \, ,
\label{qidef6}
\eeq
where $n_a \leq c$ are the number of components in the domain $D_a$.
{\color{black}
We also have that 
\beq
\mathrm{max}_{i, j \in D_a} |\langle \psi | \hat{o}_i \hat{o}_j| \psi \rangle_{\mathrm{c}}  |  \leq  \mathrm{max}_{i,j \in \ket{\psi}} |\langle \psi | \hat{o}_i\hat{o}_j| \psi \rangle_{\mathrm{c}}| \leq k_{\psi} \, ,
\eeq 
where by $i,j \in \ket{\psi}$ we denote that $i,j$ belong to the support of $\ket{\psi}$ and
the number $k_{\psi}$ is defined (if $\ket{\psi}$ is assumed normalized to 1) as the maximum difference between the squares
of the eigenvalues of $\hat{o}_i$ and $\hat{o}_j$, denoted by $\lambda_m$ and $\lambda_n$
(and $m$, $n$ labelling the eigenvalues of $\hat{o}_i$):
\beq
k_{\psi} \equiv \mathrm{max}_{\{m,n\}} \, \big( \lambda_m^2- \lambda_n^2 \big) \, .
\label{defkpsi}
\eeq 

Therefore,  from Eq. \eqref{qidef6}, and defining with the same notation of the main text
\beq	
 |V|[\ket{\psi}, \hat{O}]_{N} \equiv 4  \sum_{i,j} | \langle \psi | \hat{o}_i\hat{o}_j| \psi \rangle_{\mathrm{c}} |  = 4 \sum_{D_a} \sum_{i,j \in D_a} | \langle \psi | \hat{o}_i\hat{o}_j| \psi \rangle_{\mathrm{c}} |
 \label{defmod}
 \eeq 
 $\big(\text{the last equality holding for the same reason as for Eq. \eqref{qidef5}}\big)$, 
 we finally obtain:
\beq
F[\ket{\psi}, \hat{O}]_{N} \leq |V|[\ket{\psi}, \hat{O}]_{N} \leq 4 \, k_{\psi} \sum_{D_a} \, n_a^2  =  4 \, k_{\psi} \sum_{a = 1}^{h} \, n_a^2\, .
\label{qidef7}
\eeq
The finite number $k_{\psi}$ defined  in Eq. \eqref{defkpsi} coincides with that for  $k$ in the main text (see after Eq. (5)
)
at least for operators such that $n= -m$ (notation of the main text), as spin operators.
  
The bound in Eq. \eqref{qidef7} can be improved further, choosing  an element $D_{a^*}$ (existing by hypothesis) such that $n_{a^*} = c$: this assumption leads  to
\beq
F[\ket{\psi}, \hat{O}]_{N} \leq |V|[\ket{\psi}, \hat{O}]_{N}  \leq 4 \, k_{\psi} \, c \sum_{D_a} \, n_a  = 4 \,  k_{\psi} \,  c \, N \, ,
\label{qidef8}
\eeq
that is the bound in Eq. (8)
of the main text. From Eq. \eqref{qidef8},  it is clear that the Heisenberg scaling, $F[\ket{\psi}, \hat{O}]_{N} \sim N^2$, can strictly hold only if the parameter $k_{\psi}$ in the expressions above does not scale with $N$. This is the case of spin-$\frac{1}{2}$ operators, for which $k_{\psi} = \frac{1}{4}$. 

Importantly, also for our purposes in the Section III E
of the main text, the same bound in Eq. \eqref{boundsh} turns out to hold for the quantity $|V|[\ket{\psi}, \hat{O}]_{N} \geq F[\ket{\psi}, \hat{O}]_{N}$,
defined in Eq. \eqref{defmod} for a pure state $|\psi\rangle$, and involving the sum of modula of the connected correlations.

The expression in Eq. \eqref{qidef7} is also the starting point to prove the bound in Eq. (8)
of the main text, as shown in  Appendix A of \cite{ren2021}.
Therefore,  the same bound turns out also valid again for  $|V|[\ket{\psi}, \hat{O}]_{N}$.

We notice that in some cases (see e.g. \cite{pezze2018,gabbrielli2018}), the
fact that the connected two-point functions spatially change sign in $ F[\ket{\psi}, \hat{O}]_{N}$, $\hat{O} = \sum_i \hat{o}_i$,
can be dealt with calculating instead $ F[\ket{\psi}, \hat{O}^{\prime}]_{N}$,   $\hat{O}^{\prime} = \sum_i (-1)^i \, \hat{o}_i$. Indeed, $ F[\ket{\psi}$, where $\hat{O}^{\prime}]_{N}$ turns out equivalent to the sum of modula $ |V|[\ket{\psi}, \hat{O}]_{N}$. Therefore, in these situations it is trivially true that $ |V|[\ket{\psi}, \hat{O}]_{N}$, still being a quantum Fisher information, 
is subject to the bound in Eq. \eqref{qidef8}.

In the light of our analysis above, we stress again the deep connection between entanglement and connected two-point correlations. Importantly, it emerges quite straightforwardly
that $c$-partite can hold even if not signaled by $ F[\ket{\psi}, \hat{O}]_{N}$, for some or any choice of $\hat{O}$, as found in previous works (see e.g. \cite{pezze2017,pezze2018}).

Finally, let us discuss a bit more in detail the paradigmatic family of cases involving spin-$\frac{1}{2}$ operators.
Having in mind an array of $N$ $\frac{1}{2}$-spins, consider for instance $\hat{o}_i = s_i^{(z)}=\frac{\sigma_i^{(x)}}{2}$ \big(with $\hat{O} = \sum_i s_i^{(z)} \equiv \hat{S}^{(z)} $\big). If a pure state $|\psi \rangle$ of this system is fully producible, then $c=1$ and Eq. \eqref{qidef5} reduces to
\beq
F[\ket{\psi}, \hat{O}]_{N} =4 \, \sum_{i}  \Big[\langle \psi | s_i^{(z) \, 2}| \psi \rangle -     \langle \psi | s_i^{(z)}| \psi \rangle^2 \Big] \equiv 4 \sum_{i}  \langle \psi | s_i^{(z) \, 2} | \psi \rangle_{\mathrm{c}}  \, .
\label{qidef9}
\eeq
Therefore,  Eq. \eqref{qidef6} becomes:
\beq
F[\ket{\psi}, \hat{O}]_{N} \leq  4 |\sum_{i}  \langle \psi | s_i^{(z) \, 2} | \psi \rangle_{\mathrm{c}} | \leq  4 \sum_{i}  |\langle \psi | s_i^{(z) \, 2} | \psi \rangle_{\mathrm{c}} |  \leq  4  \,  \mathrm{max}_{i}   |\langle \psi | s_i^{(z) \, 2} | \psi \rangle_{\mathrm{c}} | \, \sum_{i}  1  \, ,
\label{qidef6sing}
\eeq
since the domain $D_a \equiv D_i$ contain now a single site (the i-th one by convention). Actually, since
\beq
 \sum_{i}  \langle \psi | s_i^{(z) \, 2} | \psi \rangle_{\mathrm{c}} = \sum_i \,  \langle \psi | \Big(s_i^{(z)} - \bar{s}^{(z)}\Big)^2 | \psi \rangle \geq 0 \, 
\eeq
(being $\bar{s}^{(z)} \equiv \frac{1}{N} \, \sum_j \, \langle \psi | s_j^{(z)}| \psi \rangle$ the average of $s_j^{(z)}$ on the $N$ sites),
then also:
\beq
F[\ket{\psi}, \hat{O}]_{N} =  4 |\sum_{i}  \langle \psi | s_i^{(z) \, 2} | \psi \rangle_{\mathrm{c}} | = 4 \sum_{i}  |\langle \psi | s_i^{(z) \, 2} | \psi \rangle_{\mathrm{c}} |  \leq  4  \,  \mathrm{max}_{i}   |\langle \psi | s_i^{(z) \, 2} | \psi \rangle_{\mathrm{c}} | \, \sum_{i}  1  \, .
\label{qidef6sing}
\eeq
Clearly $\mathrm{max}_{i}   |\langle \psi | s_i^{(z) \, 2} | \psi \rangle_{\mathrm{c}} | \leq k_{\psi} = \frac{1}{4}$, then
\beq
F[\ket{\psi}, \hat{O}]_{N}  = 4 \sum_{i}  |\langle \psi | s_i^{(z) \, 2} | \psi \rangle_{\mathrm{c}} |  \leq 4 \, k_{\psi} \, \sum_{i}  1 = 4 \, k_{\psi} \, N = N \, ,
\label{qidef10}
\eeq
correctly establishing the shot-noise limit bound for  $F[\ket{\psi}, \hat{O}]_{N}$, as well for the sum of modula $|V|[\ket{\psi}, \hat{S}^{(z)}]_{N} = 4 \sum_{i}  |\langle \psi | s_i^{(z) \, 2} | \psi \rangle_{\mathrm{c}} |$ (equal to $F[\ket{\psi}, \hat{S}^{(z)}]_{N}$ in the particular case of a fully producible state $|\psi\rangle$).}

\section*{SM 3: change of decomposition in Eq. (16) of the main text}

In this appendix, we illustrate the derivation of the general law for a change of decomposition of a density matrix $\rho$, written in Eq. (16)  of the main text.

Let us consider a generic $N \times N$ density matrix $\rho$, expressed in a certain basis (complete set) $\{\ket{v_i}\}$ on the considered $N$-dimensional Hilbert space. This
will be assumed as the working basis in all the following.

The density matrix $\rho$, being Hermitian, can be diagonalized, to $\rho_D$, via a unitary matrix $U$, by virtue of the spectral theorem of linear algebra (see e. g. \cite{lang1987}):
\beq
\rho = U \, \rho_D \, U^{-1}, \quad \quad \quad \rho_D = \mathrm{diag} (\lambda_1, \dots, \lambda_N) \, ,
\label{transf}
\eeq
and $\sum_{i =1}^N \lambda_i = 1$. 
Moreover, $\rho_D$ can be also written as a sum of projectors
\beq
\rho = U \, \rho_D \, U^{-1}, \quad \quad \quad \rho_D = \sum_{i= 1}^N \, \lambda_i \, |e_i \rangle \langle e_i| \, ,
\label{transf2}
\eeq
with 
$\{\ket{e_i}\}$  forming an orthonormal $N \times N$ set of states, still complete on the considered Hilbert space. This is called the spectral decomposition for $\rho$.
If we assume at the beginning that the set $\{\ket{v_i}\}$ was also orthonormal, then $\{\ket{e_i}\}$ and $\{\ket{v_i}\}$ can be considered connected by the unitary matrix $U$. However, the latter requirement is not central for our discussion.
Finally, we assumed, without any loss of generality,  that $\lambda_i >0, \, \,  \forall \, i$, strictly holds, exceptions being also discussed below.

From Eq. \eqref{transf} and due to the linearity of $U$, we also have:
\beq
\rho = \sum_{i =1}^N \, \lambda_i \, |\tilde{\psi}_i \rangle \langle \tilde{\psi}_i| \, ,  \quad \quad \quad  \ket{\tilde{\psi}_i} = U \ket{e_i} = \sum_{j = 1}^N \, U_{ij} \, \ket{e_j} \, ,
\label{transf3}
\eeq
where $\langle e_j | U |e_i\rangle \equiv U_{ij}$.
Therefore, the set $\{\ket{\tilde{\psi}_i}\}$ is a new decomposition for $\rho$, again orthonormal (since $U$, connecting $\{\ket{\tilde{\psi}_i}\}$ and $\{\ket{e_i}\}$, is unitary). 
Nevertheless,  $\rho$ in Eq. \eqref{transf3} is not diagonal (as it must be, being the same $\rho$ as in the left term of Eq. \eqref{transf}), because it is still expressed in the working basis $\{\ket{v_i}\}$,
 via the states $\{\ket{\tilde{\psi}_i}\}$.
 An example of this situation is the following: consider the density matrix $\rho = \frac{3}{4} \, \ket{+} \bra{+} \, + \,  \frac{1}{4} \, \ket{-} \bra{-}$, where the states $\frac{\ket{\uparrow} \pm \ket{\downarrow}}{\sqrt{2}} \equiv \ket{\pm}$ are orthonormal each other. Nevertheless, if expressed in the basis $\ket{\uparrow}$ and  $\ket{\downarrow}$, 
\beq
 \rho =  \Big(
 \begin{array}{cc}
 1/2 & 1/4\\
  1/4 & 1/2 
 \end{array}
  \Big)
 \eeq
is not diagonal.

Referring to Eq. \eqref{transf3}, different set of weights, $\{p_i\}$, can be obtained, rescaling the transformation above  by $\sqrt{p_i}$ and $\sqrt{\lambda_j}$, as follows:
\beq
 \ket{\psi_i} =  \frac{1}{\sqrt{p_i}} \,  \sum_{j =1}^N \,   U_{ij} \, \sqrt{\lambda_j} \, \ket{e_j} \, ,   \quad \mathrm{leading \, \,  to} \quad        \rho = \sum_{i =1}^N \, p_i \, |\psi_i \rangle \langle \psi_i| \, ,
\label{north}
\eeq
$\rho$ again not diagonal, in general.
After these rescalings, $\{\ket{\psi_i}\}$ are not orthogonal any longer in general, even if $\{\ket{\tilde{\psi}_i}\}$ were.  The transformation in Eq. \eqref{north} is the main result in \cite{zik}. 

If we did not assume $\lambda_i >0, \, \,  \forall \, i$, instead letting $\lambda_i = 0$, for $M <N$ eigenvalues, all the argument above can be repeated in a reduced $(N-M)$ dimensional Hilbert space, which $\rho$ can also be restricted to.\\

Now, starting from the relation 
$\ket{\psi_i} =  \frac{1}{\sqrt{p_i}} \,  \sum_{j =1}^N \,   U_{ij} \, \sqrt{\lambda_j} \, \ket{e_j}$ in Eq. \eqref{north},
changing between the spectral decomposition $\{\ket{e_i}\}$ and a generic one, $\{\ket{\psi_i}\}$, we derive how to interpolate between 
two generic decompositions $\{\ket{\psi_i}\}$, as in  Eq. \eqref{north}, and $\{\ket{\eta_i} =  \frac{1}{\sqrt{\alpha_i}} \,  \sum_{j =1}^N \,   V_{ij} \, \sqrt{\lambda_j} \, \ket{e_j}\}$ \big($\rho = \sum_{i =1}^N \, \alpha_i \, |\eta_i \rangle \langle \eta_i|$\big). Inverting the second term of  Eq. \eqref{north}, we obtain:
\beq
\ket{e_i} =  \frac{1}{\sqrt{\lambda_i}} \,  \sum_{j =1}^N \,   [U^{-1}]_{ij} \, \sqrt{p_j} \, \ket{\psi_j} \, .
\label{rchange2}
\eeq
Therefore, we can also write:
\beq
\ket{\eta_i} =   \frac{1}{\sqrt{\alpha_i}} \,  \sum_{j,k = 1}^N \,   V_{ij} \,      [U^{-1}]_{jk} \, \sqrt{p_k} \, \ket{\psi_k} = \frac{1}{\sqrt{\alpha_i}} \,  \sum_{k =1}^N \,  M_{ik} \, \sqrt{p_k} \, \ket{\psi_k} \, ,
\label{rchange3}
\eeq
$M = V \, U^{-1}$ still being a unitary matrix, since it is  the product of the unitary matrices $V$ and $U^{-1}$. Therefore, Eq. \eqref{rchange3} rules a general change of decomposition. Notice that, in Eq. (16) of the main text, we set $M \equiv U$.
}

\section*{SM 4: on Eq. (22)
on the main text and on the cluster-decomposition theorem for pure quantum states}
\label{clustersub}
In this Section, we provide additional details on the role of Eq. (22)
of the main text,
\beq
\bra{\lambda} \hat{o} (x) \hat{o} (y) \ket{\lambda} \to \bra{\lambda} \hat{o} (x) \ket{\lambda} \bra{\lambda} \hat{o} (y) \ket{\lambda}  \quad \mathrm{as}  \quad |x-y| \to \infty \, ,
\label{cdecsm}
\eeq
for a pure state $\ket{\lambda}$ on a lattice and for the specific operator $\hat{o}(x)$, as well as that of the related stronger property, called {\it cluster decomposition theorem}.  In particular, the latter theorem states that Eq. \eqref{cdecsm} holds for every local operator, and it encodes the locality principle \cite{hastings2004,sims2006,weinberg1}  for Hamiltonian systems. The same theorem can be valid at least for all the physical systems where the area law for the entanglement entropy is not violated more than logarithmically, like in systems with volume-law violation.  Eq. \eqref{cdecsm} is equivalent to the limit $\bra{\lambda} \hat{o} (x) \hat{o} (y) \ket{\lambda}_{\mathrm{c}}  \to 0$ as $|x-y| \to \infty$.

We assume at first that the pure state $\ket{\lambda}$ is not degenerate, for instance in energy.
Eq. \eqref{cdecsm} can be understood starting from
the following decomposition of  the two-points connected correlations:
\beq
\bra{\lambda} \hat{o} (x) \hat{o} (y) \ket{\lambda}_{\mathrm{c}}  = \sum_{\lambda^{\prime}}  \,  \big( \langle \lambda | \hat{o}(x) \ket{\lambda^{\prime}} \bra{\lambda^{\prime}} \hat{o}(y)| \lambda \rangle \big) - \langle \lambda | \hat{o}(x) \ket{\lambda} \bra{\lambda} \hat{o}(y)| \lambda \rangle  =  \sum_{\lambda^{\prime} \neq \lambda}  \,  \langle \lambda | \hat{o}(x) \ket{\lambda^{\prime}} \bra{\lambda^{\prime}} \hat{o}(y)| \lambda \rangle \, ,
\label{decomp}
\eeq
where $\ket{\lambda}$ and $\ket{\lambda}^{\prime}$ form again a orthonormal basis. 
In Eq. \eqref{decomp},  $ \bra{\lambda^{\prime}} \hat{o} (x)  \ket{\lambda } \to 0$ \big($ \bra{\lambda^{\prime}} \hat{o} (y)  \ket{\lambda } \to 0$\big) if $\lambda \neq \lambda^{\prime}$, then $\bra{\lambda} \lambda^{\prime} \rangle = 0$, 
 and if $x$ ($y$, or both) is (are) located on the (infinite) boundary of the system, as consequence of locality.
 The condition $ \bra{\lambda^{\prime}} \hat{o} (x)  \ket{\lambda } \to 0$ is violated by highly-entangled states  $\ket{\lambda }$,
 $\ket{\lambda^{\prime}}$, e. g. cat-states.
More  general, Eq. \eqref{cdecsm} implies that
\beq
 \sum_{\lambda^{\prime} \neq \lambda}  \, a_{\lambda , \lambda^{\prime}} \, \langle \lambda | \hat{o}(x) \ket{\lambda^{\prime}} \bra{\lambda^{\prime}} \hat{o}(y)| \lambda \rangle \to 0  \quad  \mathrm{if} \quad |x-y| \to \infty \, ,
 \label{cluster}
\eeq
at least if $|a_{\lambda , \lambda^{\prime}}| < \infty$, 
 a  generally fulfilled condition in physical systems.

The latter argument suggests that Eq. \eqref{cdecsm} and the cluster-decomposition theorem, {\color{black} often 
referred to non-degenerate ground-state(s) of local Hamiltonians, must hold also for excited states, although they are more entangled in general} (for instance  fulfilling a volume law for the Von Neumann entropy, see e. g. \cite{eisert2010,alba2009,sierra2011}). Indeed, Eq. \eqref{cluster} means that a local operator $\hat{o}(x)$ applied on (one point of) the infinite boundary of the states $\ket{\lambda}$ or $\ket{\lambda^{\prime}}$ (no matter their energies), cannot change the same states in a way to induce a nonvanishing overlap with other states of the same orthogonal basis. 

Expressed in a alternative manner, Eq. \eqref{cdecsm} and the cluster decomposition theorem must be valid for all the states of the considered systems at least if $c < \infty$, and therefore if some producibility holds. Otherwise, the entanglement between the infinite boundary and a point in the bulk, or between two points of the boundary, would immediately set $c \to \infty$. More in detail, if $N \sim L^d$, then in the latter case, $c \sim L^{\delta}$, with $\delta \geq d-1$, $d-1$ being the dimension of the boundary, then also $c \sim N^{\gamma}$, $\gamma \geq {(d-1)}/{d}$. The bound for $\delta$ can be further improved by noticing that, if Eq. \eqref{cdecsm} fails and translational invariance is assumed (at least asymptotically), then:
\beq
\langle \psi | \hat{o} (x) \hat{o} (y) | \psi \rangle_{\mathrm{c}}   \sim  |x-y|^{\beta} \, \quad \quad   \mathrm{for} \, \quad \quad   |x-y|  \to \infty \, , \quad \quad  \beta \geq 0 \, .
\label{limbound}
\eeq

Merging  Eq. \eqref{limbound} and the scaling property $f[\ket{\psi}, \hat{O}]_N \sim c \sim N^{\frac{d + 2{\beta}}{d}}$ (see more details in~\cite{hauke2016} for critical lattice systems and in \cite{pezze2017} 
for one-dimensional gapped lattices), we obtain $\delta = d +  2 \beta \geq d$.
Therefore, the failure of Eq. \eqref{cdecsm} and implies the reach of genuine multipartite entanglement. Thus, no producibility of any kind is allowed. 
The same result can be also inferred directly from the bound in Eq. \eqref{FQineqsupp}. For this purpose, it is crucial that, in the original demonstration of the bound \cite{PezzePRL2009,pezze2014}, Eq. \eqref{cdecsm} and has not been used.
For the same reason, we point out that Eq. \eqref{cdecsm}
 turns out to be strictly required to prove the bound in SM 2,
even though in those proofs no explicit mention to it is made.
Finally, the validity of Eq. \eqref{cdecsm} and of the cluster decomposition theorem does not require translational invariance, avoiding any prescription on the limit operations in Eqs. \eqref{cluster} and \eqref{limbound}.

Above, we claimed that Eq. \eqref{cdecsm} and the cluster-decomposition can hold if the ground state is unique. Instead, if the ground-state is degenerate, the same properties do not hold in a generic basis for the ground-states.  However, at least in the presence of a finite number of degenerate states, the same properties can be recovered, provided one chooses  the combination of them properly. This fact can be illustrated via a paradigmatic example, i.e. the ferromagnetic quantum Ising model in a transverse field, governed  by the Hamiltonian
\beq
\hat H_{\mathrm{IS}} = h \, \sum_{i=1}^N \hat{\sigma}_i^{(x)} + J \,  \sum_{i=1}^N \, \hat{\sigma}_i^{(z)} \hat{\sigma}_{i+1}^{(z)} \, ,  \quad   \quad J < 0  \, .
\label{ham1}
\eeq  
It is known that, if $|J| > h$, the ground state in the thermodynamic limit  is doubly degenerate, in the states $\ket{\uparrow, \dots , \uparrow} \equiv \ket{\uparrow}$ and  $\ket{\downarrow, \dots , \downarrow} \equiv \ket{\downarrow}$, orthonormal to each other. However, at finite volume, these states are mixed, forming the almost degenerate states $\ket{\pm} = \frac{\ket{\uparrow} \pm \ket{\downarrow}}{\sqrt{2}}$, exactly degenerate in the thermodynamic limit only.
For the latter states, Eq. \eqref{cdecsm} is not fulfilled, exactly since they result from the mixing of $\ket{\uparrow}$ and $\ket{\downarrow}$. Indeed:
\beq
\mathrm{lim}_{|x-y| \to \infty}  \langle \pm | \hat{\sigma}^{(z)} (x) \hat{\sigma}^{(z)} (y) | \pm \rangle \to \frac{1}{2} \quad \mathrm{but} \quad \langle \pm | \hat{\sigma}^{(z)} (x) \ket{\pm} \bra{\pm} \hat{\sigma}^{(z)} (y) | \pm \rangle = 0 \, .
\eeq
However, it is fulfilled for the states $\ket{\uparrow}$ and $\ket{\downarrow}$. This means that, adopting the basis formed by $\ket{\uparrow}$, $\ket{\downarrow}$, and by the excited states above them, Eq. \eqref{cdecsm} can be recovered, and locality made explicit.  This fact parallels the possibility, described in the main text, to hidden producibility in the presence of degeneracies. In a sense, in the presence of a finite degenerate ground state, Eq. \eqref{cdecsm} still holds, up to global unitary transformations, as that linking the states $\ket{\pm}$ and $\ket{\uparrow}$ and $\ket{\downarrow}$ in the example above.  In Eq. \eqref{cluster}, they transform the states $\ket{\lambda}$ and $\ket{\lambda}^{\prime}$ but, critically, not the operators $\hat{o}(x)$ and $\hat{o}(y)$,  set at the beginning to act on them. Therefore, the value of the sums in the same equation can change. Clearly, the bounds after Eq. \eqref{limbound} still hold; in particular not-validity of Eq. \eqref{cdecsm} in any basis implies the simultaneous absence of any sort of producibility. 

In real experiments, the states corresponding to  $\ket{\uparrow}$ and $\ket{\downarrow}$ are generically selected by the fluctuations (also classical) or by perturbations suitably added, as well as in simulations: for instance, in the example above, by an additional term $h^{\prime} \, \sum_{i=1}^N \hat{\sigma}_i^{(z)}$, $h^{\prime} \to 0$, customary e.g. in DMRG calculations.

The discussion above does not cover the important cases of continuous degeneracies, as for spontaneously broken continuous symmetries and for genuine topologically ordered matter \cite{wen2012}. Moreover, it is not clear whether a  volume-law dependence for the Von Neumann entropy is enough to guarantee the violation of Eq. \eqref{consclust} and of cluster decomposition. These issues are beyond the scope of the present work.

\section*{SM 5: derivation of Eq. (21)
 of the main text}

Starting from Eq. (6)
 of the main text,
\beq
\bar{V}[\rho , \hat{O}]_N \equiv 4 \,  \sum_{\tilde{\lambda}} p_{\tilde{\lambda}}  \, \Big(\sum_{x,y} \,  \bra{\tilde{\lambda}} \hat{o} (x) \hat{o} (y) \ket{\tilde{\lambda}}_{\mathrm{c}} \Big) \, ,
\label{defC2}
\eeq
in this Section, we obtain Eq. (21) again of the main text.
This step is strictly required to operatively evaluate $\bar{V}[\rho , \hat{O}]_N$, since in general the producible basis $\ket{\tilde{\lambda}}$, fulfilling Eq. (1)
of the main text, is not known. For this purpose, it is useful 
to consider how  Eq. \eqref{defC2} evolves under a transformation to an orthonormal, and generally entangled, basis, i.e. via the Gram-Schmidt procedure corresponding to a {\it not unitary} transformation.
Setting this transformation as
\beq
\ket{\tilde{\lambda}}  = \sum_{n} \, a_{\tilde{\lambda} , n} \,  \ket{n} \, , \quad \quad  \sum_{n} \, |a_{\tilde{\lambda} , n}|^2 \neq 1 \, ,
\label{cob}
\eeq  
where $\ket{n}$ are all orthonormal,
 and  forgetting for the moment the asymptotic terms involved in the connected correlations, we can write:
\beq
\sum_{\tilde{\lambda}} p_{\tilde{\lambda}}  \, \sum_{x,y} \,  \bra{\tilde{\lambda}} \hat{o} (x) \hat{o} (y) \ket{\tilde{\lambda}} \to  \sum_{x,y} \sum_{\tilde{\lambda}} p_{\tilde{\lambda}}  \sum_{n,m}  \, a^*_{\tilde{\lambda} , m} a_{\tilde{\lambda} , n} \,   \langle m | \hat{o} (x) \hat{o} (y) | n  \rangle \equiv \sum_{x,y}   \sum_{n,m}  \, p_{n , m}  \,   \langle m  | \hat{o} (x) \hat{o} (y) | n \rangle \,  .
\label{ftildegen}
\eeq
Exploiting again the orthonormality of the $\ket{n}$ basis, it is now easy to convince ourselves that the latter expression is equal to   
\beq
\sum_{x,y} \mathrm{Tr} \big[ \rho \, \hat{o} (x) \hat{o} (y) \big] \, ,
\label{summa}
\eeq
where
\beq
\rho = \sum_{\tilde{\lambda}} p_{\tilde{\lambda}}  \sum_{n,m}  \, a_{\tilde{\lambda} , n}  a^*_{\tilde{\lambda} , m} \,   \ket{ n}  \bra{m} \equiv \sum_{n,m} \, p_{n,m} \, \ket{n} \bra{m} 
\eeq
is now expressed in a orthonormal basis. We also have 
\beq
\mathrm{Tr} \,  \rho =  \mathrm{Tr} \,  \rho_P = \sum_{\tilde{\lambda}} p_{\tilde{\lambda}}  \sum_{n}  \, |a_{\tilde{\lambda} , n}|^2 = 1 \, .
\label{invtrace}
\eeq
Importantly, the quantity in Eq. \eqref{summa} is a trace invariant  in the space of the conceivable unitary transformations from the orthonormal basis $\ket{n}$.  
Notice also that Eq. \eqref{invtrace} is explicit if a change of basis is unitary,  $\sum_{n} \, |a_{\tilde{\lambda} , n}|^2 = 1$ as referred to Eq. \eqref{cob}, since, by hypothesis, $\sum_{\tilde{\lambda}} p_{\tilde{\lambda}} = 1$.

Let us now consider the product of matrix elements
$ \langle \tilde{\lambda} | \hat{o}(x) \ket{\tilde{\lambda}^{\prime}} \bra{\tilde{\lambda}^{\prime}} \hat{o}(y)| \tilde{\lambda} \rangle$, with $\tilde{\lambda}^{\prime} \neq \tilde{\lambda}$, like those appearing in Eq. \eqref{decomp}. 
It is immediately clear that this product can be nonvanishing only if $\bra{\tilde{\lambda}_j} \tilde{\lambda}_j^{\prime} \rangle = 0$ for just a {\it single} domain $D_j$, and only if $x$ and $y$ belong exactly to $D_j$, so that they are entangled. 
Therefore, exploiting further the cluster decomposition theorem described in SM 4,  as well as translational invariance, it turns out that
the sum in Eq. \eqref{defC2}  can be expressed again as 
\beq
\bar{V}[\rho , \hat{O}]_N =  4 \, \sum_{x,y}  \sum_{\tilde{\lambda}} p_{\tilde{\lambda}} \, \langle \tilde{\lambda}  | \hat{o} (x) \hat{o} (y) | \tilde{\lambda}  \rangle_{\mathrm{c}} =  4 \,  \sum_{x,y} \,  \sum_{\tilde{\lambda}} p_{\tilde{\lambda}}  \,  \Big( \bra{\tilde{\lambda}} \hat{o} (x) \hat{o} (y) \ket{\tilde{\lambda}} - \mathrm{lim}_{|x-y| \to \infty}  \bra{\tilde{\lambda}} \hat{o} (x) \hat{o} (y) \ket{\tilde{\lambda}} \Big) \, .
\label{ftildetraceort}
\eeq
Consequently, this can be cast in a covariant-trace form:
\beq
\bar{V}[\rho , \hat{O}]_N = 4 \,  \sum_{x,y}  \sum_{\tilde{\lambda}} p_{\tilde{\lambda}} \, \langle \tilde{\lambda}  | \hat{o} (x) \hat{o} (y) | \tilde{\lambda}  \rangle_{\mathrm{c}} = 4 \, \sum_{x,y} \Big( \mathrm{Tr} \, \big[\rho  \, \hat{o} (x) \hat{o} (y) \big]  - \mathrm{lim}_{|x-y| \to \infty}  \mathrm{Tr} \, \big[ \rho  \,  \hat{o} (x) \hat{o} (y) \big] \Big) \, .
\label{ftildetraceort2}
\eeq
As a result, in a generic {\it orthonormal} basis where in general the density matrix is not diagonal, i.e.
\beq
 \rho = \sum_{n,m} \, p_{n,m} \, \ket{n} \bra{m}  \, ,
 \label{nodiagro}
\eeq
it can finally be written as:
\beq
\bar{V}[\rho , \hat{O}]_N = 4 \, \sum_{x,y} \sum_{n,m} \,  p_{n,m}  \, \Big[ \bra{m} \hat{o} (x) \hat{o} (y)   \ket{n}  - \mathrm{lim}_{|x-y| \to \infty} \bra{m} \hat{o} (x) \hat{o} (y)   \ket{n  }  \Big]\, .
 \label{ftildestorta}
\eeq
So, apparently, we end up in the "strange correlators" $ \bra{m} \hat{o} (x) \hat{o} (y)   \ket{n}$, known to be  relevant for topology probing \cite{xu2014}.

Thanks to Eq. \eqref{cdecsm}, described in the SM 4, the second addendum in Eq. \eqref{ftildetraceort}  is physically equivalent to the term from the transformation  of the product $ \langle \tilde{\lambda}  | \hat{o} (x) | \tilde{\lambda} \rangle   \langle \tilde{\lambda} | \hat{o} (y) | \tilde{\lambda} \rangle$  in Eq. \eqref{defC2}, via the Gram-Schmidt change of basis. In this way, and oppositely to the product of one-point correlations, both the terms  in Eq. \eqref{ftildetraceort} are still expressed as a trace and are covariant under changes of basis. Therefore, Eq. \eqref{cdecsm} allows to obtain the covariant expression in Eq. \eqref{ftildetraceort2} from Eq. \eqref{defC2}, an impossible  task otherwise. 

In order to be valid for all the ground and excited states of the considered system, cluster decomposition is strictly required to guarantee some sort of producibility:  $c \sim N^{\delta}$, $0 \leq \delta < 1$, when $N  \to \infty$. Therefore, the validity of (say) Eq. (1) 
implies in itself that Eq. \eqref{cdecsm} is fulfilled for {\it all} the states $\ket{\tilde{\lambda}}$ in the same equation. 
 Finally, if the ground state is degenerate in the thermodynamic limit, we recall that the orthogonal basis $\ket{n}$ must be properly chosen to make  cluster decomposition explicit, as discussed in SM 4.

Since $\bra{m} k \rangle = \delta_{m n}$ and transformations of the basis preserve the scalar products between the states, Eq. \eqref{ftildestorta} can be simplified further. Indeed, if $m \neq n$, then $ \mathrm{lim}_{|x-y| \to \infty} \bra{m} \hat{o} (x) \hat{o} (y)   \ket{n} = 0$. This fact can be somehow justified after inserting in the last expression a complete set $\ket{\lambda}$ of states, and noticing that (for instance) $ \bra{m} \hat{o} (x)  \ket{\lambda} \to 0$ if $\lambda \neq m$ (then $\bra{m} k \rangle = 0$) and if $x$ is located on the (infinite) boundary of the system.
In this manner, Eq. \eqref{ftildestorta} can be recast as follows:
\beq
\bar{V}[\rho , \hat{O}]_N = 4 \,  \sum_{x,y}  \Big[ \sum_{n,m} \,  p_{n,m}  \,  \bra{m} \hat{o} (x) \hat{o} (y)   \ket{n}  -  \sum_{n} \,  p_{n , n} \, \mathrm{lim}_{|x-y| \to \infty} \bra{n} \hat{o} (x)   \hat{o} (y)   \ket{n} \Big]  \, .
 \label{ftildestorta2}
\eeq
 Notably, Eqs. \eqref{ftildestorta} and \eqref{ftildestorta2} become particularly manageable in the orthonormal basis $\ket{\alpha}$ where $\rho$ is diagonal. Indeed:
\beq
\bar{V}[\rho , \hat{O}]_N = 4 \,  \sum_{x,y}  \Big[ \sum_{\alpha} \,  p_{\alpha , \alpha}  \,  \bra{\alpha} \hat{o} (x) \hat{o} (y)   \ket{\alpha}  - \sum_{\alpha} \,  p_{\alpha , \alpha} \, \mathrm{lim}_{|x-y| \to \infty} \bra{\alpha} \hat{o} (x)   \hat{o} (y)   \ket{\alpha} \Big]  \, .
 \label{ftildestorta2bis}
\eeq  
The limits in Eqs. \eqref{ftildestorta}, \eqref{ftildestorta2}, and \eqref{ftildestorta2bis}  can be evaluated in a number of physically interesting cases, making these expressions manageable even at an operative level.

%
%
%
%
%
%
%
%
%


\begin{thebibliography}{XX}

\bibitem{AmicoRMP2008}
L. Amico, R. Fazio, A. Osterloh, and V. Vedral, 
{\it Entanglement in many-body systems}, 
\href{https://doi.org/10.1103/RevModPhys.80.517}{Rev. Mod. Phys. {\bf 80}, 517 (2008)}.

\bibitem{PezzeRMP}
L. Pezz\`e, A. Smerzi, M. K. Oberthaler, R. Schmied, and P. Treutlein, 
{\it Quantum metrology with nonclassical states of
atomic ensembles},
\href{https://doi.org/10.1103/RevModPhys.90.035005}{Rev. Mod. Phys. {\bf 90}, 035005 (2018)}.

\bibitem{toth2014}
G. Toth and I. Apellaniz, 
{\it Quantum metrology from a quantum information science perspective}, 
\href{https://iopscience.iop.org/article/10.1088/1751-8113/47/42/424006}{J. Phys. A {\bf 47}, 424006 (2014)}.

\bibitem{pezze2014}
L. Pezz\`{e}, and A. Smerzi
{\it Quantum theory of phase estimation},
\href{https://arxiv.org/abs/1411.5164}{arXiv:1411.5164}.
Published in {\it Atom Interferometry}, Proceedings of the International School of Physics ''Enrico Fermi'', Course 188, Varenna, pag. 691. 
Edited by G. M. Tino and M. A. Kasevich (IOS Press, Amsterdam, 2014). 


\bibitem{maccone2011}
 V. Giovannetti, S. Lloyd, and L. Maccone, 
 {\it Advances in quantum metrology}, 
 \href{https://www.nature.com/articles/nphoton.2011.35}{Nat. Phot. {\bf 5}, 222 (2011)}.

\bibitem{toth2009}
O. Guhne and G. Toth, 
{\it Entanglement detection}, 
\href{https://doi.org/10.1016/j.physrep.2009.02.004}{Phys. Rep. {\bf 474}, 1 (2009).}

\bibitem{huber2019}
N. Friis, G. Vitagliano, M. Malik, and M. Huber, 
{\it Entanglement certification from theory to experiment},
\href{https://www.nature.com/articles/s42254-018-0003-5}{Nat. Rev. Phys. {\bf 1}, 72 (2019)}.

\bibitem{HyllusPRA2012}
P. Hyllus, W. Laskowski, R. Krischek, C. Schwemmer, W. Wieczorek, H. Weinfurter, L. Pezz\`e, and A. Smerzi,
{\it Fisher information and multiparticle entanglement}, 
\href{https://doi.org/10.1103/PhysRevA.85.022321}{Phys. Rev. A {\bf 85}, 022321 (2012)}.

\bibitem{TothPRA2012}
G. T\'oth, 
{\it Multipartite entanglement and high-precision metrology}, 
\href{https://doi.org/10.1103/PhysRevA.85.022322}{Phys. Rev. A {\bf 85}, 022322 (2012)}.

\bibitem{smerzi2018}
M. Gessner, L. Pezz\`e, and A. Smerzi, 
{\it Sensitivity Bounds for Multiparameter Quantum Metrology},
\href{https://journals.aps.org/prl/abstract/10.1103/PhysRevLett.121.130503}{Phys. Rev. Lett. {\bf 121}, 130503 (2018)}.

\bibitem{HorodeckiRMP2009}
R. Horodecki, P. Horodecki, M. Horodecki, and K. Horodecki, 
{\it Quantum entanglement},
\href{https://doi.org/10.1103/RevModPhys.81.865}{Rev. Mod. Phys. {\bf 81}, 865 (2009)}.


\bibitem{Zeng}
B. Zeng, X. Chen, D.-L. Zhou, and X.-G. Wen,
{\it Quantum Information Meets Quantum Matter},
\href{https://link.springer.com/book/10.1007/978-1-4939-9084-9}{Springer (2019)}.\\
\href{https://arxiv.org/abs/1508.02595}{arXiv:1508.02595}.

\bibitem{eisert2010}
J. Eisert, M. Cramer, and M. Plenio, 
{\it Area laws for the entanglement entropy}, 
\href{https://doi.org/10.1103/RevModPhys.82.277}{Rev. Mod. Phys. {\bf 82}, 277 (2010)}.


\bibitem{OsbornePRA2002}
T. J. Osborne and M. A. Nielsen, 
{\it Entanglement in a simple quantum phase transition}, 
\href{https://doi.org/10.1103/PhysRevA.66.032110}{Phys. Rev. A {\bf 66}, 032110 (2002)}.

\bibitem{VidalPRL2003}
G. Vidal, J. I. Latorre, E. Rico, and A. Kitaev, 
{\it Entanglement in Quantum Critical Phenomena}, 
\href{https://doi.org/10.1103/PhysRevLett.90.227902}{Phys. Rev. Lett. {\bf 90}, 227902 (2003)}.

\bibitem{LatorreJPA2009}
J. I. Latorre and A. Riera,
{\it A short review on entanglement in quantum spin systems}, 
\href{https://iopscience.iop.org/article/10.1088/1751-8113/42/50/504002}{J. Phys. A {\bf 42}, 504002 (2009)}.

\bibitem{LiPRL2008}
H. Li and F. D. M. Haldane, 
{\it Entanglement Spectrum as a Generalization of Entanglement Entropy: Identification of Topological Order in Non-Abelian Fractional Quantum Hall Effect States},
\href{https://doi.org/10.1103/PhysRevLett.101.010504}{Phys. Rev. Lett. {\bf 101}, 010504 (2008)}. 

\bibitem{PollmannPRB2010} 
F. Pollmann, A. M. Turner, E. Berg, and M. Oshikawa, 
{\it Entanglement spectrum of a topological phase in one dimension},
\href{https://doi.org/10.1103/PhysRevB.81.064439}{Phys. Rev. B {\bf 81}, 064439 (2010)}.

\bibitem{FidkowskiPRL2010} 
L. Fidkowski, 
{\it Entanglement Spectrum of Topological Insulators and Superconductors}, 
\href{https://doi.org/10.1103/PhysRevLett.104.130502}{Phys. Rev. Lett. {\bf 104}, 130502 (2010)}.

\bibitem{ThomalePRL2010} 
R. Thomale, D. P. Arovas, and B. A. Bernevig, 
{\it Nonlocal Order in Gapless Systems: Entanglement Spectrum in Spin Chains}, 
\href{https://doi.org/10.1103/PhysRevLett.105.116805}{Phys. Rev. Lett. {\bf 105}, 116805 (2010)}.

 \bibitem{lepori2012}
L.~Lepori, G.~De Chiara, and A.~Sanpera, 
{\it Scaling of the entanglement spectrum near quantum phase transitions},
\href{https://doi.org/10.1103/PhysRevB.87.235107}{Phys. Rev. B {\bf 87} 235107 (2013)}.

 \bibitem{leporiLR}
 L. Lepori and L. Dell'Anna, 
 {\it Long-range topological insulators and weakened bulk-boundary correspondence}, 
\href{https://iopscience.iop.org/article/10.1088/1367-2630/aa84d0}{ New J. Phys. {\bf 19}, 103030 (2017)}.

\bibitem{OsterlohNATURE2002}
A. Osterloh, L. Amico, G. Falci, and R. Fazio, 
{\it Scaling of entanglement close to a quantum phase transition},
\href{https://doi.org/10.1038/416608a}{Nature {\bf 416}, 608 (2002)}.

\bibitem{LidarPRL2004}
L.-A. Wu, M. S. Sarandy, and D. A. Lidar,
{\it Quantum Phase Transitions and Bipartite Entanglement},
\href{https://doi.org/10.1103/PhysRevLett.93.250404}{Phys. Rev. Lett. {\bf 93}, 250404 (2004)}.

\bibitem{MartyPRL2016}
O. Marty, M. Cramer and M. B. Plenio, 
{\it Practical Entanglement Estimation for Spin-Systems Quantum Simulators},
\href{https://doi.org/10.1103/PhysRevLett.116.105301}{Phys. Rev. Lett. {\bf 116}, 105301 (2016)}.

\bibitem{PezzePRL2009}
L. Pezz\`e and A. Smerzi, 
{\it Entanglement, nonlinear dynamics, and the Heisenberg limit}, 
\href{https://doi.org/10.1103/PhysRevLett.102.100401}{Phys. Rev. Lett. {\bf 102}, 100401 (2009)}.

\bibitem{ren2021}
Z. Ren, W. Li, A. Smerzi, and M. Gessner,
{\it Metrological Detection of Multipartite Entanglement from Young Diagrams},
\href{https://journals.aps.org/prl/abstract/10.1103/PhysRevLett.126.080502}{Phys. Rev. Lett. {\bf 126}, 080502 (2021).}

\bibitem{gessner2016}
M. Gessner, L. Pezz\`e, and A. Smerzi,
{\it Efficient entanglement criteria for discrete, continuous, and hybrid variables},
\href{https://journals.aps.org/pra/abstract/10.1103/PhysRevA.94.020101}{Phys. Rev. A 94, 020101(R)  (2016)}.

\bibitem{WinelandPRA1994}
D. J. Wineland, J. J. Bollinger, M. W. Itano, and D. J. Heinzen, 
{\it Squeezed atomic states and projection noise in spectroscopy}, 
\href{https://doi.org/10.1103/PhysRevA.50.67}{Phys. Rev. A {\bf 50}, 67 (1994)}.

\bibitem{MaPHYSREP2011}
J. Ma, X. Wang, C. P. Sun, and F. Nori, 
{\it Quantum spin squeezing}, 
\href{https://doi.org/10.1016/j.physrep.2011.08.003}{Phys. Rep. {\bf 509}, 89 (2011)}.


\bibitem{sorensen2000}
A. Sorensen, L.-M. Duan, I. Cirac, and P. Zoller,
{\it Many-particle entanglement with Bose--Einstein condensates},
\href{https://www.nature.com/articles/35051038}{Nature 409, {\bf 63} (2001)}.


\bibitem{molmer2001}
A. S. Sorensen and K. Molmer,
{\it Entanglement and Extreme Spin Squeezing},
\href{https://journals.aps.org/prl/abstract/10.1103/PhysRevLett.86.4431}{Phys. Rev. Lett. {\bf 86}, 4431 (2001)}.


\bibitem{hauke2016}
P. Hauke, M. Heyl, L. Tagliacozzo, and P. Zoller, 
{\it Measuring multipartite entanglement through dynamic susceptibilities},
\href{https://doi.org/10.1038/nphys3700}{Nat. Phys. {\bf 12}, 778 (2016)}.

\bibitem{PezzePNAS2016}
L. Pezz\`e, Y. Li, W. Li, and A. Smerzi, 
{\it Witnessing entanglement without entanglement witness operators},
\href{https://doi.org/10.1073/pnas.1603346113}{PNAS {\bf 113}, 11459 (2016)}.

\bibitem{StrobelSCIENCE2014}
H. Strobel, W. Muessel, D. Linnemann, T. Zibold, D. B. Hume, L. Pezz\`e, A. Smerzi, and M. K. Oberthaler, 
{\it Fisher information and entanglement of non-Gaussian spin states},
\href{https://science.sciencemag.org/content/345/6195/424}{Science {\bf 345}, 424 (2014)}.


\bibitem{pezze2018}
M. Gabbrielli, L. Lepori, and L. Pezz\`e,
{\it Multipartite-Entanglement Tomography of a Quantum Simulator},
\href{https://iopscience.iop.org/article/10.1088/1367-2630/aafb8c}{New J. Phys. {\bf 21}  033039 (2019)}.

\bibitem{mirkhalaf2020}
S. S. Mirkhalaf, E. Witkowska, and L. Lepori,
{\it Supersensitive quantum sensor based on criticality in an antiferromagnetic spinor condensate},
\href{https://journals.aps.org/pra/abstract/10.1103/PhysRevA.101.043609}{Phys. Rev. A {\bf 101}, 043609 (2020)}.

\bibitem{MaPRA2009}
J. Ma and X. Wang, 
{\it Fisher information and spin squeezing in the Lipkin-Meshkov-Glick model},
\href{https://doi.org/10.1103/PhysRevA.80.012318}{Phys. Rev. A {\bf 80}, 012318 (2009)}.

\bibitem{LiuJPA2013}
W.-F. Liu, J. Ma, and X. Wang, 
{\it Quantum Fisher Information and spin squeezing in the ground state of the XY model},
\href{https://iopscience.iop.org/article/10.1088/1751-8113/46/4/045302}{J. Phys. A {\bf 46}, 045302 (2013)}.

\bibitem{ZhangPRL2018}
Y.-R. Zhang, Y. Zeng, H. Fan, J. Q. You, and F. Nori,
{\it Characterization of topological states via dual multipartite entanglement},
 \href{https://doi.org/10.1103/PhysRevLett.120.250501}{Phys. Rev. Lett. {\bf 120}, 250501 (2018)}.

 \bibitem{pezze2017}
L. Pezz\'e, M. Gabbrielli,  L. Lepori, and A. Smerzi, 
{\it Multipartite entanglement in topological quantum phases},
\href{https://journals.aps.org/prl/abstract/10.1103/PhysRevLett.119.250401}{Phys. Rev. Lett. {\bf 119}, 250401 (2017)}.

 \bibitem{gabbrielli2018}
M. Gabbrielli, A. Smerzi, and L. Pezz\`e,
{\it Multipartite Entanglement at Finite Temperature},
\href{https://doi.org/10.1038/s41598-018-31761-3}{Sc. Rep. {\bf 8}, 15663 (2018)}.


\bibitem{lucchesi2019}
L. Lucchesi and M. L. Chiofalo,
{\it Many-Body Entanglement in Short-Range Interacting Fermi Gases for Metrology},
\href{https://journals.aps.org/prl/abstract/10.1103/PhysRevLett.123.060406}{Phys. Rev. Lett. {\bf 123}, 060406 (2019).}

\bibitem{GuhneNJP2005}
O. G\"uhne, G. T\`oth, and H. J. Briegel, 
{\it Multipartite entanglement in spin chains},
\href{https://iopscience.iop.org/article/10.1088/1367-2630/7/1/229}{New J. Phys. {\bf 7}, 229 (2005)}.

\bibitem{daley2020}
A. Venegas-Gomez, J. Schachenmayer, A. S. Buyskikh, 
W. Ketterle, M. L. Chiofalo, and A. J. Daley,
{\it Adiabatic preparation of entangled, magnetically ordered states with cold bosons in optical lattices},
\href{https://iopscience.iop.org/article/10.1088/2058-9565/abb004}{Quant. Sc. Tech. {\bf 5} (4), 045013 (2020)}.

\bibitem{HofmannPRB2014}
M. Hofmann, A. Osterloh, and O. G\"uhne,
{\it Scaling of genuine multiparticle entanglement close to a quantum phase transition},
\href{https://doi.org/10.1103/PhysRevB.89.134101}{Phys. Rev. B {\bf 89}, 134101 (2014)}.

\bibitem{calabrese2004}
P. Calabrese and J. Cardy,
{\it Entanglement Entropy and quantum field theory},
\href{https://iopscience.iop.org/article/10.1088/1742-5468/2004/06/P06002}{J. Stat. Mech. P06002 (2004)}.

\bibitem{roscilde2018}
 I. Fr\`erot and T. Roscilde, 
 {\it Quantum Critical Metrology},
\href{https://doi.org/10.1103/PhysRevLett.121.020402}{Phys. Rev. Lett. {\bf 121}, 020402 (2018)}.
 
\bibitem{roscilde2019}
I. Fr\`erot and T. Roscilde,
{\it Reconstructing the quantum critical fan of strongly correlated systems using quantum correlations},
\href{https://doi.org/10.1038/s41467-019-08324-9}{Nat. Comm. {\bf 10}, 577 (2019)}.

\bibitem{koffel2012}
T. Koffel, M. Lewenstein, and L. Tagliacozzo, 
{\it Entanglement Entropy for the Long-Range Ising Chain in a Transverse Field},
\href{https://doi.org/10.1103/PhysRevLett.109.267203}{Phys. Rev. Lett. {\bf 109}, 267203 (2012)}.

\bibitem{ares2015}
F. Ares, J. G. Esteve, F. Falceto and A. R. de Queiroz, 
{\it Entanglement in fermionic chains with finite-range coupling and broken symmetries},
\href{https://doi.org/10.1103/PhysRevA.92.042334}{Phys. Rev. A {\bf 92}, 042334 (2015)}.

\bibitem{vodola2016}
D. Vodola, L. Lepori, E. Ercolessi, and G. Pupillo,
{\it Long-range Ising and Kitaev Models: Phases, Correlations and Edge Modes},
\href{https://iopscience.iop.org/article/10.1088/1367-2630/18/1/015001}{New J. Phys. {\bf 18}, 015001 (2016)}.

\bibitem{lepori2016}
L. Lepori, D. Vodola, G. Pupillo, G. Gori, and A. Trombettoni, 
{\it Effective theories and breakdown of conformal symmetry in a long-range quantum chain},
\href{https://doi.org/10.1016/j.aop.2016.07.026}{Ann. Phys. {\bf 374}, 35-66 (2016)}.

\bibitem{lepori2018}
 L. Lepori, D. Giuliano, and S. Paganelli, 
 {\it Edge insulating topological phases in a two-dimensional long-range superconductor},
\href{https://doi.org/10.1103/PhysRevB.97.041109}{Phys. Rev. B {\bf 97}, 041109(R) (2018)}.

\bibitem{preskill2006}
A. Kitaev and J. Preskill, 
{\it Topological entanglement entropy},
\href{https://doi.org/10.1103/PhysRevLett.96.110404}{Phys. Rev. Lett. {\bf 96}, 110404 (2006)}.

\bibitem{orus2014}
R. Orus, T.-C. Wei, O. Buerschaper, and M. Van den Nest,
{\it Geometric entanglement in topologically ordered states},
\href{https://iopscience.iop.org/article/10.1088/1367-2630/16/1/013015}{New J. Phys. {\bf 16}  013015 (2014)}.


\bibitem{nori2021}
Y.-R. Zhang, Y. Zeng, T. Liu, H. Fan, J. Q. You, and F. Nori,
{\it Multipartite entanglement of the topologically ordered state in a perturbed toric code}
\href{https://journals.aps.org/prresearch/abstract/10.1103/PhysRevResearch.4.023144}{Phys. Rev. Res. {\bf 4}, 023144 (2022)}.


\bibitem{notec}
 Note that $c$ can also diverge with $N$, $c \sim N^{l}$, $0 \leq l \leq 1$.
 
\bibitem{noteconv}
For the  spin operators,  
\color{black} $ k =  4$.  Therefore, by convention,  a normalization factor $4$ (so that $4 k  = 1$) is generally included in the definition of the QFI, 
see e.g.  \cite{pezze2014}. 


\bibitem{song2015}
 Y. Hong, S. Luo, and H. Song, 
 {\it Detecting $k$-nonseparability via quantum Fisher information},
\href{https://journals.aps.org/pra/abstract/10.1103/PhysRevA.91.042313}{Phys. Rev. A {\bf 91}, 042313 (2015)}.


\bibitem{shi2004}
Y. Shi, 
{\it Quantum Entanglement in Second-quantized Condensed Matter Systems},
\href{https://iopscience.iop.org/article/10.1088/0305-4470/37/26/014}{J. Phys. A {\bf 37}, 6807 (2004)}.

\bibitem{shi2003}
Y. Shi,
{\it Quantum Disentanglement in Long-range Orders and Spontaneous Symmetry Breaking},
\href{https://doi.org/10.1016/S0375-9601(03)00128-2}{Phys. Lett. A {\bf 309}, 254-261 (2003)}.


\bibitem{notek}
If $k$ depends explicitly on $\ket{\tilde{\lambda}}$, then in Eq. \eqref{finalboundmix},
$k = \mathrm{max}_{\tilde{\lambda}} \, k_{\tilde{\lambda}}$.


\bibitem{notasomma}
Since, for every change of decomposition, the following rule, written also in Eq. \eqref{cambiodec},
holds:  $\sqrt{p_{\lambda}} \, \ket{\lambda} = \mathrm{U}_{\lambda \, \lambda^{\prime}} \, \sqrt{p_{\lambda^{\prime}}} \,  \ket{\lambda^{\prime}}$,
as derivable from \cite{zik} {\color{black} (see SM 3)}. Therefore, $ \sum_{\lambda} \, p_{\lambda} = \mathrm{Tr} \, \big[\rho \, {\bf I} \big]$, as immediate
if $\ket{\lambda}$ is orthonormal, is invariant under changes of decompositions by the unitaries ${\bf U}$ (between the states rescaled by $\sqrt{p_{\lambda}}$).   
In the ultimate analysis, the same result stems from the completeness of each decomposition.


\bibitem{yu2013}
S. Yu,
{\it Quantum Fisher Information as the convex roof of variance},
\href{https://arxiv.org/abs/1302.5311}{arXiv:1302.5311}.


\bibitem{petz2013}
G. Toth and D. Petz,
{\it Extremal properties of the variance and the quantum Fisher information},
\href{https://journals.aps.org/pra/abstract/10.1103/PhysRevA.87.032324}{Phys. Rev. A {\bf 87}, 032324 (2013)}.


\bibitem{note1}
Therefore, apart from the contribution coming from
$\bar{V}_1[\rho , \hat{O}]_N$, the QFI is obtained by the maximization
on the $\bar{V}_2[\rho , \hat{O}]_N$'s evaluated
on all the decompositions $\ket{\lambda}$,
obtained as explained above
from the particular decomposition  $\ket{\tilde{\lambda}}$ entering
Eq. \eqref{cpartmix31}.


\bibitem{lindblad1976}
G. Lindblad,
{\it  On the generators of quantum dynamical semigroups},
\href{https://doi.org/10.1007/BF01608499}{Comm. Math. Phys. {\bf 48}, 119  (1976)}.

\bibitem{gorini1976}
V. Gorini, A. Kossakowski, and E. C. G. Sudarshan,
{\it  Completely positive dynamical semigroups of N-level systems},
\href{https://doi.org/10.1063/1.522979}{J. Math. Phys. {\bf 17}, 821 (1976)}.

\bibitem{breuerpetruccione2002}
H.-P. Breuer and F. Petruccione,
{\it The theory of open quantum systems},
Oxford University Press, Oxford (2002).



\bibitem{weinberg1}
S. Weinberg,
{\it The Quantum Theory of Fields: Foundations},
Cambridge University Press, Cambridge, UK (1995), vol. 1.

\bibitem{hastings2004}
M. B. Hastings, 
{\it Locality in Quantum and Markov Dynamics on Lattices and Networks},
\href{https://doi.org/10.1103/PhysRevLett.93.140402}{Phys. Rev. Lett. {\bf 93}, 140402 (2004)}.

\bibitem{sims2006}
B. Nachtergaele and  R. Sims, 
{\it Lieb-Robinson Bounds and the Exponential Clustering Theorem},
\href{https://doi.org/10.1007/s00220-006-1556-1}{Comm. Math. Phys. {\bf 265}, 119 (2006)}.

\bibitem{gp}
G. Grosso and G. Pastori Parravicini,
{\it Solid State Physics}, 
Elsevier (2014).

{\color{black}
\bibitem{lepori2022}
L. Lepori, M. Burrello, A. Trombettoni, and S. Paganelli,
{\it Strange correlators for topological quantum systems from bulk-boundary correspondence},
\href{https://journals.aps.org/prb/abstract/10.1103/PhysRevB.108.035110}{Phys. Rev. B {\bf 108}, 035110 (2023)}.
}

\bibitem{gessner2017_2}
M. Gessner, L. Pezz\`e, and A. Smerzi,
{\it Entanglement and squeezing in continuous-variable systems},
\href{http://quantum-journal.org/papers/q-2017-07-14-17/}{Quantum {\bf 1}, 17 (2017)}.

\bibitem{zik}
I. Bengtsson and K. Zyczkowski,
{\it Geometry of quantum states (An Introduction to Quantum Entanglement)},
Cambridge University Press, New York, (2006).\\

\vspace{1cm}

\bibitem{uhlmann1976}
A. Uhlmann,
{\it The "transition probability" in the state space of a $*$-algebra},
\href{https://www.sciencedirect.com/science/article/abs/pii/0034487776900604}{Rep. Math. Phys. {\bf 9}, 273-279 (1976)}.

\bibitem{wang2014}
J. Liu, X.-X. Jing, W. Zhong, and X. Wang,
{\it Quantum Fisher information for density matrices with arbitrary ranks},
\href{https://iopscience.iop.org/article/10.1088/0253-6102/61/1/08}{Comm. Theor. Phys., {\bf 61}(01), 45-50 (2014)}.

\bibitem{lang1987}
S. Lang,
{\it Linear algebra},
Undergraduate Texts in Mathematics,
Springer (1987).


\bibitem{alba2009}
V. Alba, M. Fagotti, and P. Calabrese,
{\it Entanglement entropy of excited states},
\href{https://iopscience.iop.org/article/10.1088/1742-5468/2009/10/P10020}{J. Stat. Mech. P10020 (2009)}.
 
 \bibitem{sierra2011}
 M. Ibanez Berganza, F. Castilho Alcaraz, and G. Sierra,
 {\it Entanglement of excited states in critical spin chains},
 \href{https://iopscience.iop.org/article/10.1088/1742-5468/2012/01/P01016}{J. Stat. Mech. P10016 (2011)}. 
 
 \bibitem{wen2012}
X. G. Wen,
{\it Topological order: from long-range entangled quantum matter to an unification of light and electrons},
\href{https://www.hindawi.com/journals/isrn/2013/198710/}{ISRN Cond. Matt. Phys., 2013 198710 (2013)}.

\bibitem{xu2014}
Y.-Z. You, Z. Bi, A. Rasmussen, K. Slagle, and C. Xu,
{\it Wave Function and Strange Correlator of Short-Range Entangled States},
\href{https://doi.org/10.1103/PhysRevLett.112.247202}{Phys. Rev. Lett. {\bf 112}, 247202 (2014)}.


\end{thebibliography}
\end{document}